\documentclass[]{aa}

\usepackage{txfonts}			

\usepackage[utf8]{inputenc}	
\usepackage{amsmath}			
\usepackage{nccmath}			

\usepackage[english]{babel}	
\usepackage{xcolor}			
\usepackage{graphicx}			
\usepackage{booktabs}			
\usepackage{gensymb}			
\usepackage{subcaption}

\usepackage[
colorlinks = true,
anchorcolor = blue,
linkcolor = blue,
urlcolor  = blue,
citecolor = blue,
]{hyperref}

\usepackage{textcomp}
\usepackage{orcidlink}
\usepackage{multicol}
\usepackage{multirow}
\usepackage{lscape}
\usepackage{tipa}
\usepackage{physics}
\newcommand{\lhood}{$\mathcal{L}(\mathcal{D}|\theta)$\xspace}
\newcommand{\prior}{$P(\theta)$\xspace}
\newcommand{\posterior}{$P(\theta | \mathcal{D})$\xspace}

\newcommand{\micron}{\textmu{}m}

\let\oldsim\sim
\renewcommand{\sim}{{\oldsim}}

\newcommand{\achlys}{Achlys\xspace}
\newcommand{\achlyslong}{(208996) Achlys\xspace}

\newcommand{\gkun}{G!k\'un\textdoublepipe\textprimstress h\`omd\'im\`a\xspace}
\newcommand{\gkunlong}{(229762) \gkun}

\newcommand{\huya}{Huya\xspace}
\newcommand{\huyalong}{(38628) Huya\xspace}

\newcommand{\varda}{Varda\xspace}
\newcommand{\vardalong}{(174567) Varda\xspace}

\newcommand{\tc}{84522\xspace}
\newcommand{\tclong}{(84522) 2002~TC302\xspace}

\newcommand{\kx}{119951\xspace}
\newcommand{\kxlong}{(119951) 2002~KX14\xspace}

\newcommand{\mani}{M\'ani\xspace}
\newcommand{\manilong}{(307261) M\'ani\xspace}

\newcommand{\vs}{84922\xspace}
\newcommand{\vslong}{(84922) 2003~VS2\xspace}

\newcommand{\ixion}{Ixion\xspace}
\newcommand{\ixionlong}{(28978) Ixion\xspace}

\newcommand{\oc}{470316\xspace}
\newcommand{\oclong}{(470316) 2007~OC10\xspace}

\begin{document}

\title{Probing close-in satellites of Trans-Neptunian Objects through thermal and direct size measurements} 
\titlerunning{Probing close-in satellites of Trans-Neptunian Objects through thermal and direct size measurements}

%
%
\author{
J.~M.~Gómez-Limón\inst{1}\orcidlink{0009-0006-8584-1416}\and
R.~Leiva\inst{1}{\orcidlink{0000-0002-6477-1360}} \and
J.~L.~Ortiz\inst{1}\orcidlink{0000-0002-8690-2413} \and
P.~Santos-Sanz\inst{1}\orcidlink{0000-0002-1123-983X} \and
M.~Kretlow\inst{2,1}\orcidlink{0000-0001-8858-3420} \and
Y.~Kilic\inst{1}\orcidlink{0000-0001-8641-0796} \and
J.~L.~Rizos\inst{1}\orcidlink{0000-0002-9789-1203} \and
A.~Álvarez-Candal\inst{1}\orcidlink{0000-0002-5045-9675} \and
T.~Müller\orcidlink{0000-0002-0717-0462} \inst{3}
}

\authorrunning{Gómez-Limón et al.}

\institute{
Instituto de Astrofísica de Andalucía, IAA-CSIC, Glorieta de la Astronomía s/n, 18008 Granada, Spain. \and Deutsches Zentrum für Astrophysik (DZA), Postplatz 1, 02826 Görlitz, Germany. \and Max-Planck-Institut für extraterrestrische Physik, Giessenbachstr. 1, 85748 Garching, Germany. 
} 

\date{Received dd mmm, 2026; accepted dd mmm, 2026}

\abstract
{ Trans-Neptunian objects are distant bodies that retain valuable information about the origin and evolution of the Solar System. Many of these objects constitute binary systems. Studying binaries allows us to further characterise this primitive population and is critical for determining mass densities, a key but elusive physical property. Nevertheless, satellite detection can be challenging.}
%
{ This study aims to constrain the presence of close-in satellites around a selection of ten trans-Neptunian objects, including four known binary systems used for methodology validation.}
%
{We developed a methodology independent of primary-secondary separation. We exploit the combination of occultation-derived sizes and thermal emission data from the "TNOs are Cool" \textit{Herschel Space Observatory} key project. We model the thermal emission from a binary system to explain the thermal excess that cannot be reproduced by a single body of the occultation-derived size.}
%
{ We obtain satisfactory constraints for the validation targets \achlyslong{}, \gkunlong{}, \huyalong{} and \vardalong{}. We find that \tclong{}, \kxlong{}, and \manilong{} are likely binary systems, which was previously unknown. We report size estimates for their putative satellites. For \vslong{}, \ixionlong{}, and \oclong{} we find that no sizable satellite is needed to reconcile thermal and occultation data.}
%
{}
 

\maketitle

\section{Introduction}\label{sec:introduction}

Trans-Neptunian Objects (TNOs) populate the region of the Solar System beyond Neptune's orbit known as the Kuiper Belt. These objects are remnants of the planetesimal formation in the protoplanetary disk that gave birth to the planetary system we inhabit today, and they retain crucial information about its evolution (e.g. \cite{nesvorny2018dynamical, fernandez2020introduction, gladman2021transneptunian}).

Trans-Neptunian Binaries (TNBs) are pairs of gravitationally bound TNOs. The Pluto/Charon system was the first binary system discovered in the Kuiper belt \citep{christy1978satellite}, although we now know it is a multiple system \citep{weaver2006discovery, weaver2016small}. However, 35 years had to elapse for a second detection of a TNB to be confirmed \citep{veillet2002binary}. Since then, many new TNBs have been discovered, accounting for a total of 144 as of February 2026 \footnote{See \url{https://www.johnstonsarchive.net/astro/asteroidmoons.html}}. The study of TNBs has thrived in the last two decades, being one of the most active branches of Kuiper Belt research (e.g. \cite{grundy_mutual_2019, noll_chapter_2020, porter_detection_2024}).

One of the reasons for trying to detect binaries in the Kuiper Belt is that a characterisation of the mutual orbit constrains the system mass \citep{grundy_five_2011,parker_characterization_2011,grundy_mutual_2019}. If independent size measurements are available, the density can be retrieved, informing us about bulk properties such as ice-to-rock ratio or porosity. These properties are extremely difficult to attain otherwise.

The study of TNBs is also a pivotal method for testing the prediction of planetesimal formation theories and models. For example, the abundance of binary prograde orbits provides strong evidence in favour of the Streaming Instability (SI) \citep{nesvorny_trans-neptunian_2019}. This theory suggests that solid objects in the circumsolar disk began forming when small clusters of solid material created localised enhancements in gas drag, enabling these clusters to grow rapidly into full-sized TNOs \citep{youdin2005streaming,johansen2007rapid,bai2010dynamics,carrera2017planetesimal}.

The SI has grown to be the leading theory in planetesimal formation, and it is predicted to be very efficient in forming binary systems of roughly equal size \citep{nesvorny2010formation}. This is consistent with observations suggesting that planetesimals formed almost entirely in binary pairs \citep{fraser_all_2017}. The chances are high that some of these primordial binary pairs survived their implantation in their current orbits, especially if they are tight systems \citep{nesvorny_binary_2019}. In addition, \cite{porter_kctf_2012} performed Kozai Cycle Tidal Friction simulations, showing that TNBs orbits tend to shrink and circularise with time, leading to roughly $\sim 30 \%$ of their randomised systems evolving into tight orbits with eccentricity $e < 10^{-4}$.

Other proposed mechanisms for binary system formation in the Kuiper Belt are giant collisions \citep{canup2005giant}, rotational fission \citep{ortiz2012rotational} and grazing encounters \citep{leinhardt2010formation} or 'kiss-and-capture' scenarios \citep{denton2025capture}. In general, these mechanisms generate satellites around the parent bodies that might not be as equal-sized as those created by the SI. Most of the largest TNOs have moons. This is the case for 7 of the 8 largest TNOs (apart from Pluto): Eris \citep{holler2021eris}, Haumea \citep{ragozzine2009orbits}, Makemake \citep{parker2016discovery}, Gonggong \citep{kiss2017discovery}, Quaoar \citep{fraser2013mass}, Orcus \citep{sickafoose2019stellar}, and Salacia \citep{stansberry2012physical}, with Sedna being the exception.

It is clear that the current picture is that the Kuiper Belt should be densely populated with binary systems. The largest objects tend to be orbited by smaller moons, whereas roughly equal-sized binaries are expected at smaller sizes, with preference for tight orbits. However, the current collection of known TNBs clearly lacks these tight systems: Of the 144 known TNBs, only 5  (3.5~\%) have separations below 800 ~km, whereas 52 (36.1~\%) have separations less than 3000 ~km. The reason behind these numbers is an observational bias because most TNBs are discovered via direct imaging with the Hubble Space Telescope (HST) \citep{noll_chapter_2020}. The pixel scale of the Wide Field Camera 3 (WFC3), which is the optimal HST instrument for TNB detection, is 40 mas/pixel, equivalent to $\sim 1200$ ~km at 40 AU. With the most advanced point spread function fitting methods, binaries can be detected as close as half a WFC3 pixel ($\sim 600$ ~km at 40 AU), although this is only possible for equal-sized TNBs with a combined brightness of $V\simeq21$ . The resolution quickly degrades for fainter TNOs (see Figure 3 in \cite{porter_detection_2024}). 

HST observations are further constrained by several operational and environmental factors. Reliable target acquisition requires the availability of suitable guide stars within the limited guide-star patrol field, a condition that is not always met for TNO observations. In addition, when TNOs are projected against densely populated stellar fields (such as during passages near the Galactic plane) source confusion and background contamination might render the data unuseful. These conditions reduce sensitivity to close binaries and may therefore limit the suitability or prioritisation of such targets for HST observations.

The goal of this paper is to overcome the limitations of direct imaging in the search for TNBs. To do so, we have developed a methodology to quantitatively exploit the combination of thermal data and stellar occultation measurements. This methodology allows us to indirectly infer the presence of satellites around TNOs independently of the separation distance, as long as the system is unresolved. The basic idea is to include occultation derived sizes in the analysis of mid- and far-infrared measurements. It has been observed that the occultation diameters, which are direct measurements, tend to be slightly smaller than those obtained by fitting single-object models to thermal data. This difference can be attributed to the presence of an unresolved companion, whose flux also contributes to the infrared measurements \citep{ortiz2020thermal}. Consequently, when a single-object thermal model is fitted to these measurements, an artificially larger diameter is obtained to account for the extra thermal emission from the satellite.

The "TNOs are Cool" project \citep{muller2009tnos} was a \textit{Herschel Space Observatory} open time key programme dedicated to observing the thermal emission of TNOs and Centaurs. This programme also incorporated measurements from \textit{Spitzer}. A total of 180 objects were observed under this programme \citep{muller_chapter_2020}. Radiometric sizes and albedos have been published for most of these objects from fits of the Near Earth Asteroid Thermal Model (NEATM, \cite{harris_thermal_1998}), or slight modifications to it. It is important to note that these results come from considering a single object. In this paper, we use the colour corrected mid- and far-infrared measurements from "TNOs are Cool" publications.

Stellar occultations offer a way of determining sizes of Solar System bodies with kilometric accuracy \citep{ortiz2020bookchapter}.  For the first time, a significant sample of TNOs with a direct size measurement is available. Some of them were also targets of "TNOs are Cool". Hence, we are in conditions to combine these datasets and set the first preliminary constraints on the presence of satellites among these bodies, free of the separation bias.

This work is structured as follows:  In Section \ref{sec:methodology} we describe in detail the modelling we implemented and the Bayesian approach that is followed for parameter determination. In Section \ref{sec:sampleselection} we describe the TNOs that will be analysed with this methodology. In Section \ref{sec:observational_data} we expose the datasets used. The concrete application of the aforementioned methodology to our target selection is detailed in Section \ref{sec:Analysis} and the results are presented and discussed in Section \ref{sec:results}.

\section{Methodology}\label{sec:methodology}

In the original "TNOs are Cool" works, a single-object model is considered by default. In this paper, we use the NEATM \citep{harris_thermal_1998} to model both components of a TNB. Application of the NEATM to TNOs has been extensively discussed in many of the "TNOs are Cool" publications (e. g. \citealt{vilenius_tnos_2012}, \citealt{santos-sanz_tnos_2012}). In \cite{gomez2025size}, a binary object model consisting in two NEATM bodies was applied for the first time to model the thermal emission of a TNO in combination with occultation measurements. In this Section, we present an improved methodology based on the one presented in that paper which will be applied to the selection of TNOs specified in Section \ref{sec:sampleselection}.

\subsection{The model} \label{sec:themodel}

The NEATM assumes a spherical object in the limit of zero thermal inertia and/or no rotation. In other words, the temperature at each surface point only depends on the solar illumination  geometry. Therefore, the temperature distribution on the surface of the sphere is given by \citep{delbo_thermal_2007}
\begin{equation}
    \label{eq:temperature_distribution}
    T = \begin{cases}
    T_{SS}(\cos i)^{1/4}, & \text{if $i<\pi /2$}.\\
    0, & \text{if $i>\pi /2$}.
  \end{cases}
\end{equation}
Here $i$ is the angular distance of a surface point to the sub-solar point\footnote{Point where solar radiation is normal to the surface.} and $T_{SS}$ is the temperature at the sub-solar point. The value of the sub-solar temperature is determined by equating the total energy absorbed by a surface element to that emitted in the thermal infrared. A factor $\eta$, known as the beaming parameter, is included in the energy balance equation to account for deviations from the assumptions (like the thermal inertia not being negligible). The expression for the sub-solar temperature is then \citep{delbo_thermal_2007}
\begin{equation}
\label{eq:Tss}
    T_{SS} = \qty[\frac{(1-A)S_{\odot}}{\epsilon \eta \sigma (r_h/1\ \text{AU})^2} ]^{1/4},
\end{equation}
where $A$ is the Bond albedo, $S_\odot = 1367$~W m$^{-2}$ is the solar constant \citep{thuillier2004solar}, $\sigma$ is the Stefan-Boltzmann constant, $\epsilon$ is the emissivity, and $r_h$ is the heliocentric distance of the modelled object. Throughout this work, a grey emissivity ($\epsilon = 0.9$) is assumed, as has been the case for the publications of "TNOs are Cool" (\cite{lellouch_tnos_2013}. Since the solar spectrum peaks near the V-band, we assume $A\simeq A_V = qp_V$. Here, $q$ is the phase integral and $p_V$ is the geometric albedo (in the V-band). For consistency with the TNOs are Cool studies, we adopt the same approximation for the phase integral ($q = 0.336 p_V + 0.479$; \citealt{brucker2009high}). We note that more recent determinations of phase integrals for TNOs exist \citep{verbiscer2022diverse}, but we retain this formulation to ensure direct comparability.

The disc integrated flux is then computed in the NEATM by assuming grey-body radiation at each surface element visible to the observer. The phase angle is taken into account, although its effect is almost negligible for TNOs since they are always observed at very small phase angle from Earth's perspective. The value of this integral depends on the diameter of the sphere $D_{\text{eq}}$. Therefore, for a single object, the observed flux at a certain thermal wavelength under the NEATM is a function only of the beaming parameter $\eta$, the geometric albedo $p_V$ and the object diameter $D_{\text{eq}}$.

From the definition of absolute magnitude $H_V$ for a Solar System body, the geometric albedo is given by
\begin{equation}
    \label{eq:albedo}
    p_V = \frac{\pi a^2}{S_{\text{proj}}}10^{\frac{2}{5}(V_\odot - H_V)},
\end{equation}
where $a=1$ AU, $V_\odot = -26.76 \pm 0.02$ mag is the apparent V magnitude of the Sun \citep{bessell1998model} and $S_{\text{proj}}$ is the projected area towards the observer, which for a spherical model is $S_{\text{proj}} = \pi D_{\text{eq}}^2/4$. Using Eq. \ref{eq:albedo}, one can exchange the dependency of NEATM on $p_V$ for dependency on $H_V$. Consequently, for a single object, we have $ F_{\text{model}}(\lambda) = F_{\text{NEATM}}(H_V, D_{\text{eq}}, \eta ; \lambda)$.
 
In the present work, we assume that both components of the binary system are unresolved. This implies that measurements of the absolute magnitude represent the brightness of the combined system. We adopt a model where both components have equal surface properties, i.e., equal beaming parameter and geometric albedo. Given the diameters of the primary $D_{\text{eq}}$ and secondary $d_{\text{eq}}$ and the combined absolute magnitude $H_V$, one can compute the projected area of the combined system $S_{\text{proj}} = \pi(d_{\text{eq}}^2 + D_{\text{eq}}^2)/4$ and use Eq. \ref{eq:albedo} to retrieve the common albedo for both objects. Once $p_V$ is known, we can invert Eq. \ref{eq:albedo} to obtain the individual absolute magnitudes $H_{V,\text{main}}$ and $H_{V,\text{sat}}$. Hence, we can write for the combined system:

\begin{equation}
\label{eq:flux_binary}
\begin{split}
F_{\text{model}}(\lambda) &= F_{\text{NEATM}}\left(H_{V,\text{sat}}(H_V, D_{\text{eq}}, d_{\text{eq}}),\, d_{\text{eq}},\, \eta ;\, \lambda\right) \\
&\quad +\; F_{\text{NEATM}}\left(H_{V,\text{main}}(H_V, D_{\text{eq}}, d_{\text{eq}}),\, D_{\text{eq}},\, \eta ;\, \lambda\right).
\end{split}
\end{equation}

In the present work, we use the available direct measurements of $H_V$ and $D_{\text{eq}}$ in the literature. The absolute magnitude measurements come from regular ground-based photometric observations. For the diameter of the primary, we use occultation data. In general, the detected projected limbs are not circular, and ellipses are fitted to the observed occultation chords in the literature. To assign a diameter value to the occultation measurement, we use the area-equivalent diameter, i.e., the diameter of a circle of equal area to the fitted ellipse.

All the occultation data used in this work correspond to detections of a single object. We assume that the detected object is the largest component of the pair in case the object is actually a binary system. This is the most likely situation, since smaller satellites might lie undetected in between chords or outside the scanned area of the sky plane. However, nowhere on the previous equations is imposed that $d_{\text{eq}} < D_{\text{eq}}$, so this modelling can also be applied in the case where the detected object is the secondary/satellite (in which case $d_{\text{eq}}$ and $D_{\text{eq}}$ would exchange roles).

Since $H_V$ and $D_{\text{eq}}$ have known values, from Eq. \ref{eq:flux_binary} it is clear that we only have $\eta$ and $d_{\text{eq}}$ as free parameters to model the observed flux at certain thermal wavelength $\lambda$.

\subsection{Bayesian statistical approach} \label{sec:Bayesian}

We follow a Bayesian approach for parameter estimation. We choose this methodology because it is the most appropriate way to propagate the uncertainties in $D_{\text{eq}}$ and $H_V$ in our analysis. Bayesian parameter estimation also allows for a full characterisation of uncertainties and parameter correlation, and allows computing credible intervals or upper limits on the parameter values in a natural way.

The central goal of Bayesian inference is to characterise the posterior probability density function (posterior pdf or simply posterior). This distribution describes the conditional probability of the model parameters $\theta = (d_{\text{eq}}, \eta)$ given the measured thermal fluxes $\mathcal{D} = \{ F_{\text{meas}}(\lambda _i)\}_{i}$ and considering  our prior knowledge of the parameters. According to the Bayes Theorem, the posterior \posterior can be expressed as

\begin{equation}
\label{eq:Bayes}
    P(\theta | \mathcal{D}) = \frac{\mathcal{L}(\mathcal{D} | \theta) P(\theta)}{P(\mathcal{D})},
\end{equation}
where \lhood is the likelihood function, \prior is the prior pdf and $P(\mathcal{D})$ is known as the evidence \citep{gelman2013bayesian, robert2007bayesian}.

\prior represents our belief on the parameter values prior to the measurements whose analysis will result in an update on such beliefs. We assume that, a priori, $\eta$ and $d_{\text{eq}}$ are independent, and hence the prior pdf can be written as \prior $=f_1(\eta)f_2(d_{\text{eq}}; H_V,D_{\text{eq}})$.

The likelihood function \lhood measures how likely it is to observe the data $\mathcal{D}$ for different parameter values $\theta$. In our case, the data are the flux measurements at thermal wavelengths from "TNOs are Cool". We consider that the flux errors $\sigma _i$ are independent and normally distributed, in which case the likelihood may be written as \citep{gregory2005bayesian}:
\begin{equation}
    \label{eq:likelihood}
\mathcal{L}(\mathcal{D}|\theta) = 
(2\pi)^{-N/2}
\Biggl[ \prod_{i=1}^N \frac{1}{\sigma _i} \Biggr]
\exp\left[ -\sum_{i=1}^N 
    \frac{ \left[ F_{\text{meas}}(\lambda_i) - F_{\text{model}}(\lambda_i) \right]^2 }{2\sigma_i^2}
\right]
\end{equation}

Estimates on the value of the beaming parameter are given for all our targets in previous "TNOs are Cool" publications. However, these estimates are based on the fitting of single-object models. Consequently, we cannot use these values individually as prior information for  each of our targets, since we are considering a binary object model. Instead, we choose to use for every target the distribution of $\eta$ values across the TNO population. We take the 68 beaming parameter values fitted in \cite{lellouch_tnos_2013} (TNOs only) and estimate $f_1(\eta)$ using Gaussian Kernel Density Estimation with the \texttt{gaussian\_kde} class of \texttt{scipy.stats} Python package.

For the prior in the diameter of the secondary, we consider a uniform distribution for all targets. Some of our targets have known satellites (see Section \ref{sec:sampleselection}), and size estimates are available in the literature. However, we ignore this information in our analysis so that these objects serve as validation for our methodology. The limits of the uniform distribution $d_{\text{low}}$ and $d_{\text{high}}$ depend on the nominal values of the occultation derived diameter $D_{\text{eq}}$ and the absolute magnitude of the system $H_V$. In general, if the common geometric albedo $p_V$ is known, one can infer the value of $d_{\text{eq}}$ by computing $H_{V,\text{main}}$ from Eq. \ref{eq:albedo}, then obtaining $H_{V,\text{sat}}$ from $H_{V,\text{main}}$ and $H_V$, and finally applying Eq. \ref{eq:albedo} again. $d_{\text{low}}$ is the result of this computation if one assumes $p_V=0.01$ and $d_{\text{high}}$ is the result if $p_V = 1.15$. In practice, this does not exclude a priori any reasonable $d_{\text{eq}}$ values. However, imposing these limits is necessary to ensure no extreme $p_V$ values are internally encountered in the sampling process. Summing up, we choose $f_2(d_{\text{eq}}; H_V,D_{\text{eq}})$ to be the pdf of a uniform distribution between $d_{\text{low}}(H_V, D_{\text{eq}})$ and $d_{\text{high}}(H_V, D_{\text{eq}})$, reflecting that we have no a priori information on the size of any putative companion, except that the albedo cannot be extreme.

To approximate the posterior distribution, we employ a Markov Chain Monte Carlo (MCMC) method to generate samples from it. For this purpose, we use the Python package \texttt{emcee 3.1.6} \citep{foreman-mackey_emcee_2013}, which implements the affine invariant sampling algorithm described in \cite{goodman_ensemble_2010}.

In each of our MCMC samplings, we used \texttt{emcee} with $n_w = 128$ random walkers (i.e. parallel chains in the ensemble sampler), running the sampler for $n_{\text{burn}}=500$ initial steps. The samples from the initial steps are subsequently discarded (burn-in phase), in order to reduce the influence of the initial conditions before the chains reach their stationary distribution. We adopt the criterion $n_{\text{burn}}\ge 10 \tau$ in both $d_{\text{eq}}$ and $\eta$, where $\tau$ is auto-correlation time estimated by \texttt{emcee}. The criterion ensures that the chain is evolved long enough to effectively lose dependence of its initial state. Starting from the last position of the burn-in phase, the sampler is run for $n_{\text{iter}}$ additional iterations. The sample size from this final phase is therefore $n_w \times n_{\text{iter}}$, which is representative of the posterior distribution \posterior. 
For a comprehensive explanation of these criteria and parameters, we refer the reader to the original emcee papers and references therein.

To propagate the uncertainties in the values of $H_V$ and $D_{\text{eq}}$, which have been independently directly measured, we choose a scheme in which we perform a total of $500$ MCMC runs for each analysed object. These runs are the result of dividing the $D_{\text{eq}} \times H_V$ parameter subspace in a grid, and performing each run with a fixed value combination of $(D_{\text{eq}},H_V)$ in the generation of our modelled fluxes with Eq. \ref{eq:flux_binary}. To generate the grid, we draw $n=500000$ $(H_V,D_{\text{eq}})$ pairs from the uncertainty distributions of $H_V$ and $D_{\text{eq}}$, considering the variables are independent. In most cases, the literature values report a symmetric error bar, in which case we consider a normal distribution with $\sigma$ equal to the reported error. From this sample, we then perform a bidimensional histogram of 500 equal-sized bins; 50 bins for the case of $D_{\text{eq}}$ and 10 bins for the case of $H_V$. We have a lower resolution in the absolute magnitude dimension because $d_{\text{eq}}$ (the parameter we are most interested in) shows low dependency on $H_V$, see Section \ref{sec:General_results}.

In each 2D bin, we run the burn-in phase, and the number of additional steps $n_{\text{iter}}$ is proportional to the number of counts in the bin  $m_{\text{bin}}$: $n_{\text{iter}} = \lfloor m_{\text{bin}} / n_w \rfloor$. Then we combine all $500$ samples into a single one, incorporating the $(H_V,D_{\text{eq}})$ values for each of them. Therefore, the final sample will consist of $n$ four-dimensional vectors $(H_V,D_{\text{eq}},d_{\text{eq}},\eta)$, where only the knowledge of $d_{\text{eq}}$ and $\eta$ is being updated in this Bayesian scheme.

\section{Sample selection} \label{sec:sampleselection}

Our basic target selection criterion is that the TNO should have published thermal emission and occultation measurements. We restricted to occultations with at least three positive chords, so that a reliable projected area can be retrieved.

We explicitly excluded the remarkable cases of Quaoar and Haumea, since they are bodies known to possess rings \citep{ortiz2017size, morgado2023dense}, a feature that is not considered in our model and which would lead to systematic errors in the analysis. On top of this, these objects have already been thermophysically modelled in detail, see \cite{kiss_visible_2024} and \cite{santos-sanz_tnos_2017} respectively. Similarly, Makemake is also excluded from our analysis. It is a system with complex thermal emission \citep{kiss_prominent_2024} and for which intricate two-terrain thermal models have been applied in the past \citep{lim_tnos_2010}. Eris is also excluded because, although it technically has a three positive chord occultation, two of the chords correspond to telescopes located in the same observatory. Therefore, the area-equivalent diameter is poorly constrained, as it might vary from $\sim$2300~km from a circular fit to $\sim$ 3000 km from an elliptical fit \citep{sicardy2011pluto}.

(143707) 2003 UY117 is a "TNOs are Cool" target with a published three-chord occultation \citep{kretlow2024physical}. However, the thermal emission coverage is very poor, as it was only observed at 70 \micron{} and 160 \micron{}, and for the latter the emission was so low that only an upper limit in flux is reported \citep{farkas-takacs_tnos_2020}. Therefore, we also exclude this object from our study.



The rest of TNOs fulfilling the aforementioned criterion are included in our target selection and are listed on Table \ref{tab:merged_table}. We have a sample of ten TNOs of diverse dynamical classes, four of which are already known binary systems:
\begin{itemize}
    \item \achlyslong (provisional designation 2003~AZ84) has a satellite $5.0 \pm 0.3$ magnitudes fainter than the primary, discovered from HST \citep{brown2007satellites}. Assuming equal albedos for primary and secondary, measurements of the primary by stellar occultations deliver an approximate size for the secondary of $\sim 80$ ~km in diameter \citep{dias2017study}. The semi-major axis of the orbit is roughly 7200 ~km, corresponding to $< 300$ mas of projected separation at \achlys{} distance.

    \item \gkunlong{} (provisional designation 2007~UK126) is a system extensively studied with HST data \citep{grundy2019Gkunhondima}. Its satellite G!\`{o}'\'{e}!h\'{u} is $3.242\pm 0.039$ magnitudes fainter. The semi-major axis of the orbit is well constrained to be $a=6035 \pm 48$ ~km, equivalent to $<250$ mas at its distance.

    \item The satellite of \huyalong{} (provisional designation 2000~EB173) was discovered using HST in 2012 \citep{noll201238628}. Stellar occultations by the satellite provide a very reliable lower limit for its size, set to $d_{\text{sat}} > 165$ ~km in \cite{rommel2025stellar}. From the same work, the semi-major axis of the orbit is determined to be $a=1898^{+22}_{-22}$ ~km ($<100$ mas of separation), being the tightest known binary system in our sample.

    \item \vardalong (provisional designation 2003~MW12) has a large satellite named Ilmarë, which has also been extensively studied from HST and other large ground-based telescopes \citep{grundy_mutual_2015}. These observations allowed the fitting of the orbit, resulting in a semi-major axis of $a=4812 \pm 35$ ~km, or $<200$ mas at \varda{}'s distance. The observed magnitude difference in the system is $\Delta m = 1.734 \pm 0.042$ \citep{grundy_mutual_2015}.
    
\end{itemize}

The previous four systems serve as validation of the methodology. To first order, the presence of a satellite should be recovered in these cases, corroborating that similar satellites should also be recovered in the case of smaller separation distances. To second order, we can verify if the satellite diameter derived from our methodology is compatible with the estimations done with independent measurements.

It is important to note that, in all four cases, the maximum apparent separation from Earth is well below 0.5 arcsec. Since the average seeing from the ground is over 0.5 arcsec in the overwhelming majority of situations, the photometric observations of these targets from the ground will measure the blended flux of both the main body and the satellite (unless adaptive optics are used). In other words, the system is unresolved in ground-based observations without adaptive optics, as those used to obtain the absolute magnitude of the objects. Therefore, the values in the literature reflected in Table \ref{tab:merged_table} take into account the combined flux of both bodies. The spatial resolution of \textit{Herschel}/PACS and \textit{Spitzer}/MIPS is $\sim 5$ arcsec or worse \citep{rieke2004multiband, Poglitsch2010}. Consequently, the thermal measurements also represent the combined infrared flux of the system.

Currently, we do not have direct evidence that the remaining TNOs in our target list (\tc{}, \kx{}, \mani{}, \vs{}, \ixion{} and \oc{}) possess any satellite. However, it must be noted that these objects have been studied non-homogenously with HST. We performed a search in the Mikulski Archive for Space Telescopes (MAST \footnote{\url{https://mast.stsci.edu/search/ui/\#/}}) taking into account provisional designation, permanent designation and numbering of the objects. According to this search:

\begin{itemize}
    \item The objects \tc{}, \kx{}, \mani{}, \vs, and \ixion{} have been observed under the HST proposal ID 10545. 
    One image was taken for \tc{}, four for \ixion{}, and two for the other objects. The aim of the proposal was to discover and characterise satellites around TNOs using the Advanced Camera for Surveys (ACS) instrument. No detections were ever reported for these targets.
    
    \item \ixion{} was also observed under the HST proposal ID 9110, with a similar goal of discovering Charon-analogue satellites. \ixion{} observations consisted of two images using the Space Telescope Imaging Spectrograph (STIS) instrument. No satellite has been reported.
    
    \item We found no HST observations for \oc{}.
\end{itemize}

\section{Observational data}\label{sec:observational_data}

 A summary of the $D_{\text{eq}}$ and $H_V$ values used and the corresponding references can be found in Table \ref{tab:merged_table}. In the following, we give some details on the particularities of each target. Specifically, we report the mid- and far-infrared wavelength coverage by "TNOs are Cool" measurements and the corresponding sources. We take the flux values and uncertainties as they are. We also take from these sources the relevant geometrical parameters from the observations: target-observer distance, heliocentric distance, and phase angles. The coverage can include two bands from \textit{Spitzer}/MIPS centred at 24 \micron{} and 71 \micron{}, and three bands from \textit{Herschel}/PACS at 70 \micron{}, 100 \micron{} and 160 \micron{}. Exceptionally, some "TNOs are Cool" targets were also observed with \textit{Herschel}/SPIRE (bands at 250 \micron{}, 350 \micron{} and 500 \micron{}).

\subsection{\achlyslong{}}

The case of \achlys{} is of special interest. \citep{dias2017study} report two multi-chord occultations by this TNO, with 3 positives each. The fitted projected ellipses are considerably different, both in eccentricity and size. In \cite{dias2017study}, the authors argue that this represents a triaxial body observed at two different rotational phases. A Jacobi ellipsoid is fitted to these projections and the observed rotational light curve amplitude. A mean area-equivalent diameter of $772\pm12$ ~km is reported (rotational average) for the derived three-dimensional shape. We implement this $D_{\text{eq}}$ value in our analysis.

\achlys{} was intensively observed with \textit{Herschel}/PACS within the "TNOs are Cool" programme. The objective was to measure its rotational light curve at thermal wavelengths, although this was not achieved \citep{santos-sanz_tnos_2017}. In this work we use the averaged \textit{Herschel}/PACS values reported in Table 2 from \cite{mommert_tnos_2012}. These do not include measurements at 70 \micron{}. We also take the \textit{Spitzer}/MIPS measurements from the same publication.

\subsection{\gkunlong{}{}}

\gkun{} produced a stellar occultation in 2014 (8 positive chords) that has been analysed in \cite{schindler2017results} and in \cite{benedetti2016results}. In \cite{schindler2017results} only 3 chords are used in the analysis of the projected shape (well spaced and with reliable GPS timings), although the resulting fitted ellipse is almost identical to that reported in \cite{benedetti2016results}, which uses more chords. We decide to use as area-equivalent diameter the one from \cite{schindler2017results} because the authors used only the data with highest timing accuracy. In this work, also volume-equivalent diameter estimates are provided. However, in this case we decided not to use these values since they are informed by the same thermal data we use here, but assuming a single-object model.

Flux measurements in the three \textit{Herschel}/PACS  bands are available for this object \citep{santos-sanz_tnos_2012}. \cite{schindler2017results} revisit the originally published values, and re-reduce the data following the technique described in \cite{kiss2014optimized}. Because of the higher SNR and improved calibration using stars of similar brightness to the target, we use these revised data as our thermal dataset.

The absolute magnitude of the system is also revised in \cite{schindler2017results}. The reported value is the same as in \cite{perna2010colors}, but with larger uncertainty that takes into account effects from potential rotational variability. This $H_V$ value incorporates the flux contribution from the satellite G!\`{o}'\'{e}!h\'{u}.

\subsection{\huyalong{}{}}

\huya{} has two published multi-chord occultations, one from 2019 \citep{santos2022physical} and one from 2023 \citep{rommel2025stellar}. In both cases, the observed projected area is very similar, with area-equivalent diameters of $411.0 \pm 7.3$ ~km and  $420.6 \pm 22.0$ ~km respectively. Because of the almost unchanging observed projections, we stick with the value reported in \cite{santos2022physical} for our analysis, because the 2019 occultation has higher SNR in the occultation light curves, leading to a smaller uncertainty.

\cite{fornasier_tnos_2013} report thermal measurements for the two \textit{Spitzer}/MIPS bands, the three \textit{Herschel}/PACS bands, and also measurements at 250 \micron{}, 350 \micron{} and 500 \micron{} from \textit{Herschel}/SPIRE. ALMA observations of \huya{} at 1.29~mm \citep{lellouch2017thermal} are also included in our analysis. This object exhibits the broadest thermal wavelength coverage among the TNOs in our sample.

\subsection{\vardalong{}{}}

\cite{souami2020multi} report the only multi-chord occultation to date of \varda{}, with five positive chords. Fluxes in the three \textit{Herschel}/PACS bands for \varda{} are available in \cite{vilenius_tnos_2014}.

Two non-compatible absolute magnitudes of the Varda-Ilmarë system are reported in the literature. \cite{vilenius_tnos_2014} derive a value of $H_V = 3.61\pm 0.05$ using data from \cite{perna2013photometry}. \cite{alvarez2016absolute} estimate an absolute magnitude of $H_V= 3.998\pm 0.048$ using literature data and including data from Observatorio de Sierra Nevada 1.5 m and Calar Alto 3.5 m telescopes. The main difference between the two works is that \cite{vilenius_tnos_2014} use a fixed phase coefficient equal to the mean of the TNO population \citep{belskaya2008bookchapter}, whereas \cite{alvarez2016absolute} fit the phase coefficient to their dataset. For this reason, we believe the latter is more reliable.

\subsection{\tclong{}}

This object was successfully observed via a stellar occultation in 2018, with 12 positive detections (a record coverage at the time for a TNO besides Pluto). Its projected area at the moment of the occultation was determined with notable precision \citep{ortiz2020large}.

Within the "TNOs are Cool" programme, it was observed with \textit{Herschel}/PACS and \textit{Spitzer}/MIPS. With MIPS, the flux at 71 \micron{} could not be recovered because the object was merged with a background source, so only the flux at 24 \micron{} is available \citep{fornasier_tnos_2013}.

The absolute magnitude value we use for this object is $H_V=4.23$, reported in \cite{tegler2016two}. However, they report no uncertainty, so we assume a conservative $\sigma_{H_V}=0.1$ as they report colours for this object with $\sigma  = 0.02$  mag.

\subsection{\kxlong{}}

The object \kx{} is currently crossing the galactic plane and consequently has undergone numerous occultation events, several of which have multiple positive detections. In \cite{rizos2025trans} a collection of 4 multi-chord events is published. Their analysis concludes that all the occultations are coherent, within uncertainties, to the same projected ellipse (the three dimensional shape is likely an oblate spheroid). In our analysis, we use the area-equivalent diameter from their elliptical fit to all the occultations events.

Thermal emission wavelength coverage of \kx{} is extensive, with measurements in the three \textit{Herschel}/PACS bands and the two \textit{Spitzer}/MIPS bands \citep{vilenius_tnos_2012}.

\subsection{\manilong{}{}}

\mani{} is currently the TNO with the most positive chords ever detected in an occultation: 61. The coverage was so comprehensive that even topographic features could be identified \citep{rommel2023large}. For the present work, we take the area-equivalent diameter from the global elliptical fit to a selection of chords (those with the most accurate timings) in \cite{rommel2023large}, see their Table 4.

As was the case for \kx{}, the thermal emission wavelength coverage from \textit{Herschel}/PACS and \textit{Spitzer}/MIPS is complete \citep{vilenius_tnos_2012}.

\subsection{\vslong{}}

The case of \vs is similar to \achlys{}. It also has two published multi-chord stellar occultations, one in 2014 \citep{benedetti2019trans} and one in 2019 \citep{vara2022multichord}. Its shape likely departs from a spheroid, since it has a large rotational light curve amplitude \citep{sheppard2007light, ortiz2006short}. Because the rotational phase at the time of the 2019 occultation was exceptionally well constrained, in \cite{vara2022multichord} a mean area equivalent diameter is reported, which represents the rotational average. We use this value of $D_{\text{eq}} = 545  \pm 13$ ~km in our analysis.

As for \achlys{}, extensive observations of \vs were carried out to look for rotational variation in the thermal emission \citep{santos-sanz_tnos_2017}. As before, we consider the rotationally averaged flux values from \textit{Herschel}/PACS reported in Table 2 from \cite{mommert_tnos_2012}. In this case, the 100 \micron{} band is not available. We also take the \textit{Spitzer}/MIPS measurements from the same publication.

\subsection{\ixionlong{}{}}

The case of \ixion{} is similar to that of \kx{}, as it is also immersed in a region of the sky densely populated by stars. Therefore, several multi-chord occultations have been measured in recent years. A compilation of them has been published in \cite{Kilic2026Ixion}. In this work, the authors follow a methodology similar to \kx{} \citep{rizos2025trans}: Since the projected profile seems to be unchanged across occultations, a single ellipse is fitted to all the combined occultations, leading to the area-equivalent diameter value presented in Table \ref{tab:merged_table}, that we incorporate to our analysis.
Fluxes in the three \textit{Herschel}/PACS bands for \ixion{} were published in \cite{lellouch_tnos_2013}.

\subsection{\oclong{}}

An occultation of a double star by \oc{} in 2022 delivered 4 positive chords \citep{gomez2025size}. This object is the smallest in our sample, and also has the largest uncertainties in projected size due the moderate SNR in the occultation light curves. Because this object is the only object in our sample with an occultation analysed under a Bayesian framework, it is also the only one with asymmetrical error bars in the reported area-equivalent diameter.

Complete \textit{Herschel}/PACS coverage of this object is available in \cite{santos-sanz_tnos_2012}. This object has already been analysed using a methodology almost identical to that proposed in Section \ref{sec:methodology}, the only difference being that uncertainties in absolute magnitude were not considered \citep{gomez2025size}. Therefore, it serves as a useful test to check how our inference on putative satellite presence changes when introducing uncertainties in $H_V$.

\begin{table*}
\caption{Summary of dynamical, physical, and rotational properties for the analyzed targets.}
\label{tab:merged_table}
\renewcommand{\arraystretch}{1.3} 
\centering
\begin{tabular}{llp{1.3cm}p{2.5 cm}lp{3cm}}
\hline \hline
Object & Orbital Class & Known Binary & $D_{\text{eq}}$ (km) & $H_V$ (mag) & Rot. LC amplitude (mag) \\
\midrule
\achlyslong{} & Plutino       & Yes & $772\pm 12$ (D17) & $3.779\pm 0.114$ (A16) & $0.07 \pm 0.01$ (T10) \\
\gkunlong{}   & Detached      & Yes & $621\pm 7$ (S17)  & $3.69 \pm 0.10$ (S17)  & $0.03 \pm 0.01$ (T14) \\
\huyalong{}   & Plutino       & Yes & $411.0\pm 7.1$ (S22) & $5.04\pm 0.03$ (F13) & $0.02\pm 0.01$ (T14) \\
\vardalong{}  & Resonant 15:8 & Yes & $740\pm 14$ (S20)  & $3.998\pm 0.048$ (A16) & $0.02 \pm 0.01$ (T14) \\
\midrule
\tclong{}     & Resonant 5:2  & No  & $500\pm 10$ (O20)  & $4.23 \pm 0.10$ (T16)  & $0.06\pm 0.01$ (O20) \\
\kxlong{}     & Classical $^*$    & No  & $382.2\pm 8.7$ (R25) & $4.978\pm 0.017$ (A16) & $<0.05$ (R25) \\
\manilong{}   & Hot Classical     & No  & $796\pm 24$ (R23)  & $3.63\pm 0.05$ (R23)   & $0.05\pm 0.01$ (T13) \\
\vslong{}     & Plutino       & No  & $545\pm 13$ (V22)  & $4.14\pm 0.07$ (A16)   & $0.264\pm 0.017$ (V22) \\ 
\ixionlong{}  & Plutino       & No  & $697\pm 10$ (K26)  & $3.845\pm 0.006$ (K26) & $<0.15$ (O03) \\
\oclong{}     & Detached      & No  & $330^{+56}_{-55}$ (G25) & $5.40\pm 0.02$ (G25) & $<0.1$ (G25) \\
\bottomrule
\end{tabular}
\tablefoot{Orbital classification follows \cite{volk2024dynamical}. $^*$ Classification of \kx{} as hot or cold classical is debated \citep{fernandez2021compositional, rizos2025trans}. The values of primary area-equivalent diameter and system absolute magnitude used in our analyses are the ones listed in the  $D_{\text{eq}}$ and $H_V$ columns respectively. For thermal measurements sources, please refer to Section \ref{sec:observational_data}.}
\tablebib{
(A16) \cite{alvarez2016absolute}; (D17) \cite{dias2017study}; (F13) \cite{fornasier_tnos_2013}; 
(G25) \cite{gomez2025size}; (K26) \cite{Kilic2026Ixion}; (O03) \cite{ortiz2003study}; 
(O20) \cite{ortiz2020large}; (R23) \cite{rommel2023large}; (R25) \cite{rizos2025trans}; 
 (S17) \cite{schindler2017results}; (S20) \cite{souami2020multi}; 
(S22) \cite{santos2022physical}; (T10) \cite{thirouin2010short}; (T13) \cite{thirouin_thesis_2013}; 
(T14) \cite{thirouin2014rotational}; (T16) \cite{tegler2016two};
(V22) \cite{vara2022multichord}. 
}
\end{table*}

\section{Analysis}\label{sec:Analysis}

The parameter we are most interested in is the area-equivalent diameter of the satellite, $d_{\text{eq}}$. By analysing the marginalised posterior of $d_{\text{eq}}$ we can distinguish between two cases:
\begin{enumerate}
    \item The marginalised posterior peaks at $d_{\text{eq}} = 0$. In this case, according to our analysis, a single-object model is coherent with both thermal data and occultation observations. We can use this marginalised posterior to set quantitative upper limits on the presence of any companion. For example, the 68~\% credibility upper limit is given by the 68~\% quantile. 
    Our inference on the beaming parameter for these cases can be directly compared to the results from "TNOs are Cool" works, where a single-object model is assumed by default (in \cite{lellouch_tnos_2013}, for example). Our result will represent an updated value since "TNOs are Cool" analyses feature $D_{\text{eq}}$ as a free parameter to be fitted, whereas we adopt the diameter independently derived from occultations.
    
    \item The marginalised posterior shows a peak at $d_{\text{eq}} > 0$. In this case, the conclusion is that a binary object explains better the occultation and thermal data. We give an estimate of the size of the putative satellite by reporting the mean of the marginalised posterior in $d_{\text{eq}}$. The corresponding error bars are computed by taking the 68 \%  maximum density credible interval in this marginalised posterior. That is, the shortest interval containing 68 \% of the sample elements, corresponding to the $1\sigma$ region.
\end{enumerate}

We flag these two cases for our targets in the the second column in Table \ref{tab:results_summary}. For all targets, the nominal value of $\eta$ is reported as the median of the marginalised sample, and the error bar is computed from the 68 \%  maximum density credible interval.

\section{Results and Discussion}\label{sec:results}

\begin{figure}[h!]
\centering

\begin{subfigure}{\linewidth}
    \centering
    \includegraphics[width=\linewidth]{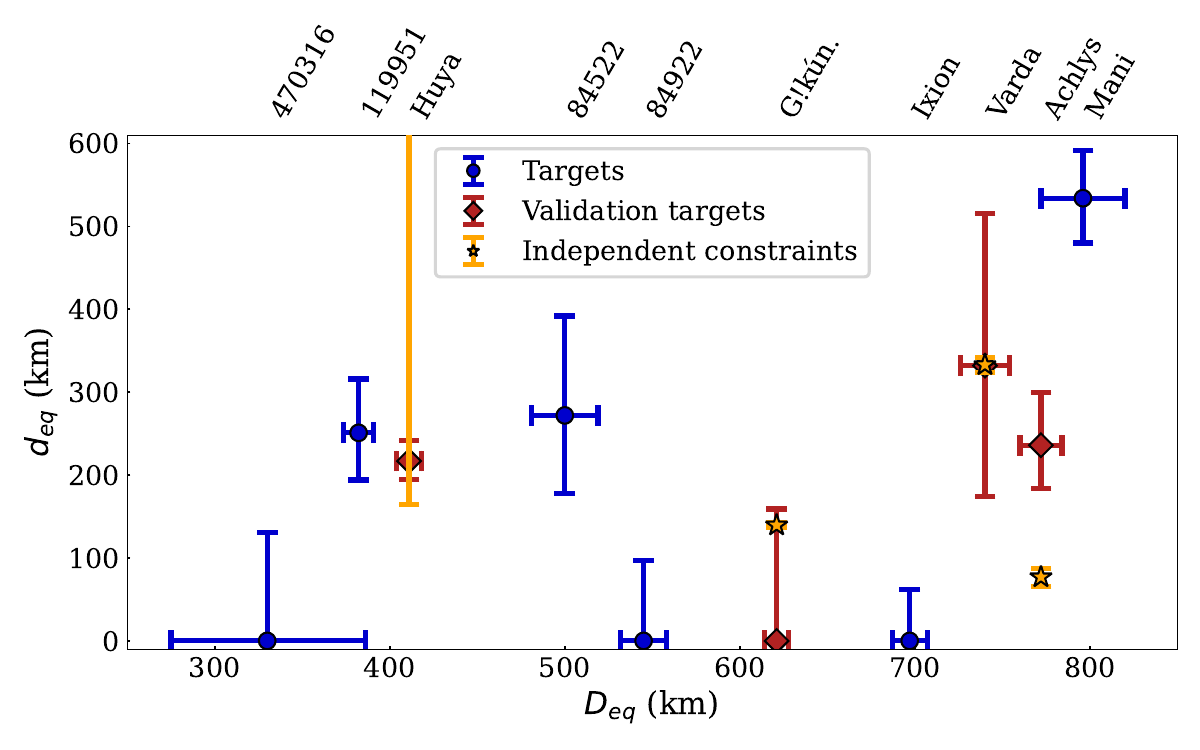}

\end{subfigure}

\begin{subfigure}{\linewidth}
    \centering
    \includegraphics[width=\linewidth]{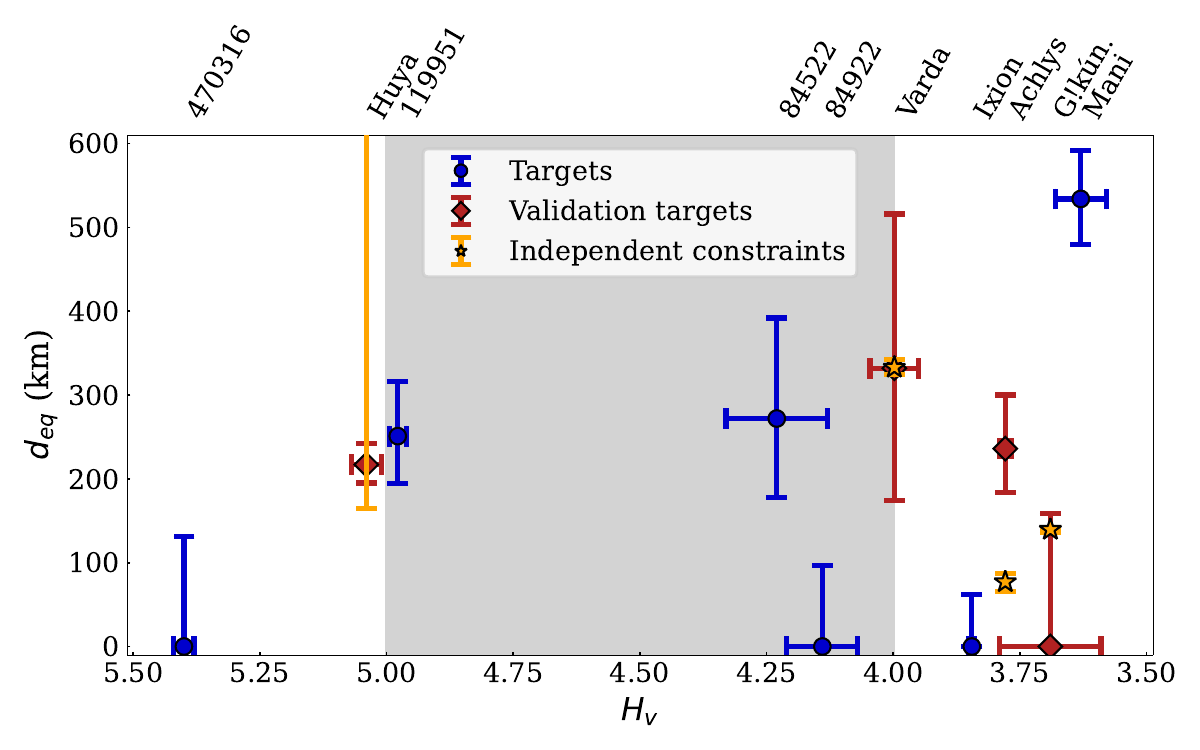}
\end{subfigure}

\caption{Inferred satellite sizes for the target sample. The upper panel displays satellite size as a function of the primary’s occultation-derived area-equivalent diameter. The lower panel shows satellite size as a function of the system absolute magnitude obtained from independent photometric observations. The interval $H_V \in [4,5]$, region where a dearth of binaries has been reported in \cite{lyra_where_2025}, is depicted in grey. For Huya, the independent upper limit on $d_{\text{eq}}$ is represented with the orange errorbar.}

\label{fig:summary}
\end{figure}

\begin{figure*}[htbp]
    \centering

    \begin{subfigure}[b]{0.42\textwidth}
        \centering
        \includegraphics[width=\linewidth,height=\linewidth]{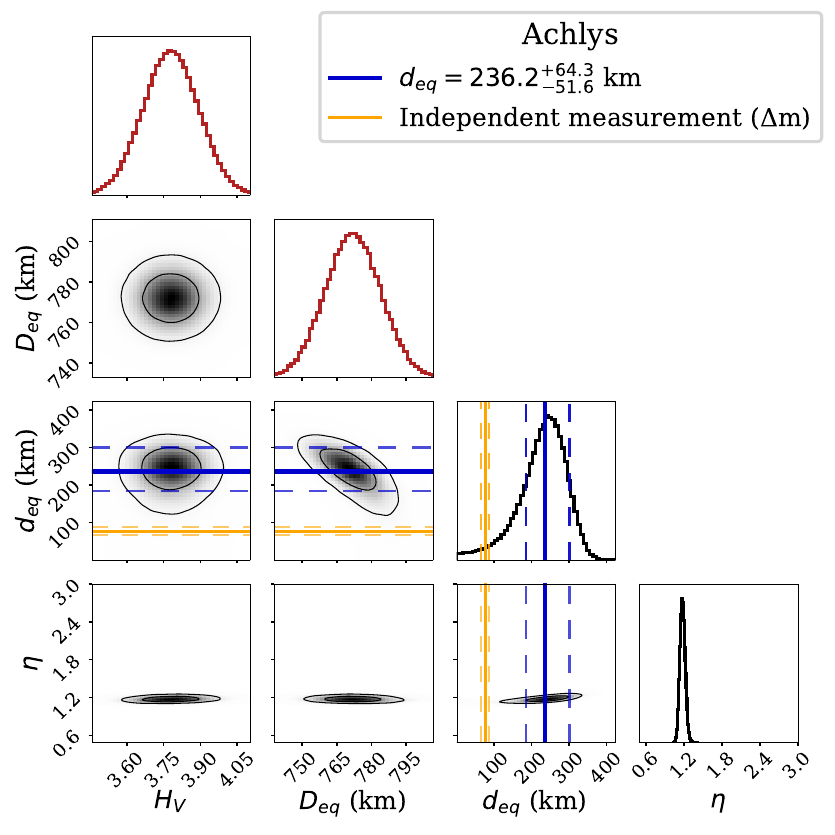}
    \end{subfigure}
    \hspace{1.5cm}
    \begin{subfigure}[b]{0.42\textwidth}
        \centering
        \includegraphics[width=\linewidth,height=\linewidth]{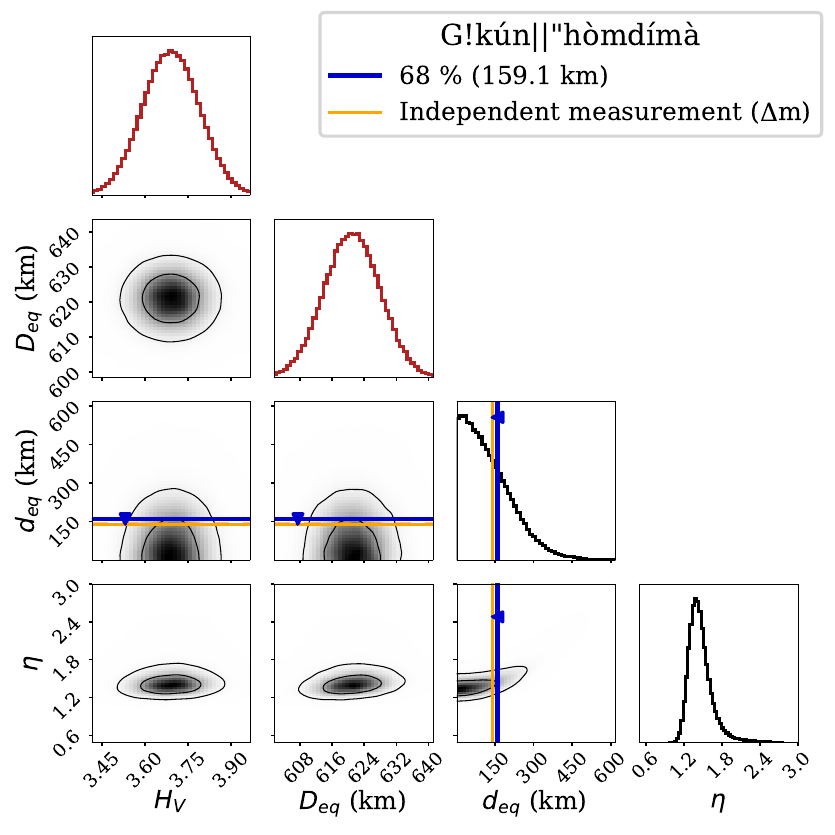}
    \end{subfigure}

    \begin{subfigure}[b]{0.42\textwidth}
        \centering
        \includegraphics[width=\linewidth,height=\linewidth]{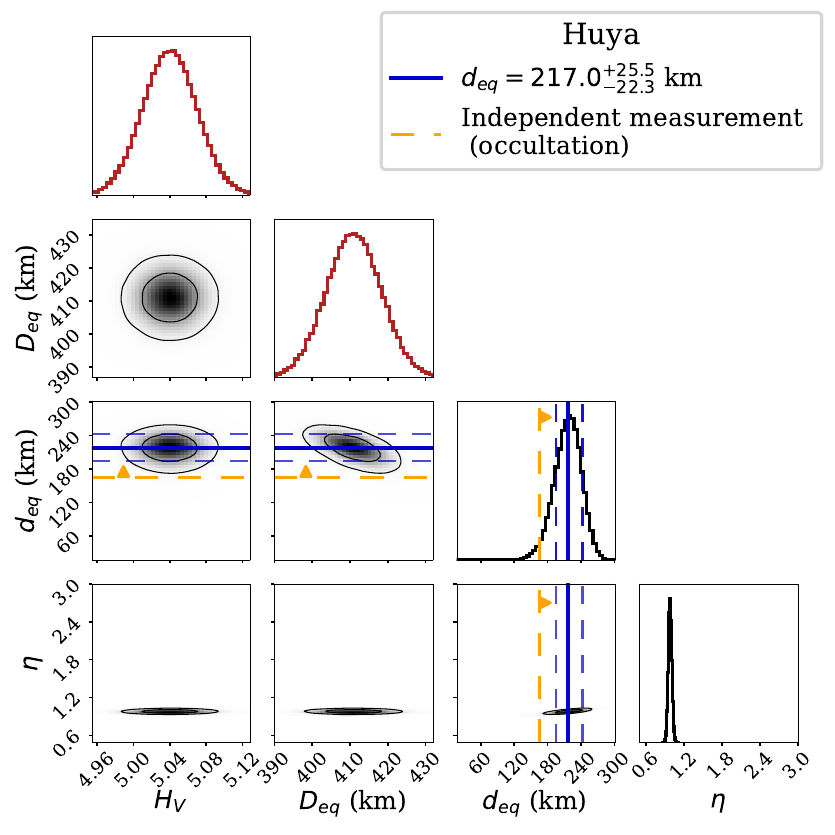}
    \end{subfigure}
    \hspace{1.5cm}
    \begin{subfigure}[b]{0.42\textwidth}
        \centering
        \includegraphics[width=\linewidth,height=\linewidth]{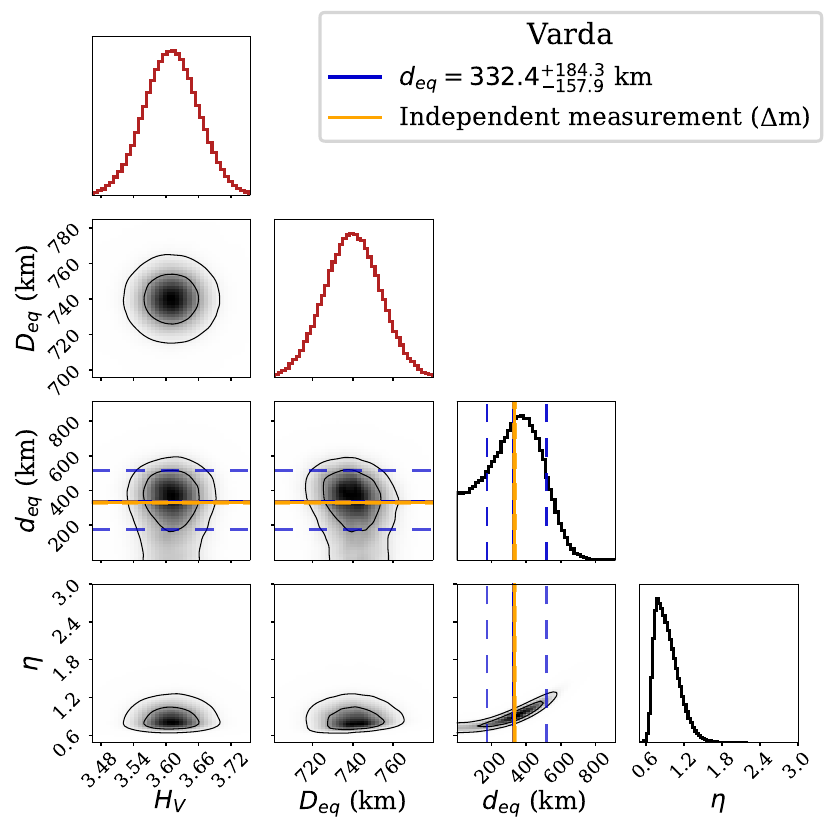}
    \end{subfigure}
    
    \caption{Pairwise plots for the validation targets. The marginalised posterior pdfs of $H_V$ and $D_{\text{eq}}$ are shown in red to highlight the fact that these are not free parameters. These pdfs represent the independent measurements of $H_V$ and $D_{\text{eq}}$ from literature, reported in Table \ref{tab:merged_table}, that were included in the analysis. The constraints on the size of the satellite from independent measurements are overplotted in yellow, see Table \ref{tab:results_summary}. Except for Huya, these independent constraints come from the observed magnitude difference, which we translate into a secondary area-equivalent diameters under the assumption of equal albedos. The propagated uncertainties for these diameters (dashed yellow lines) are small. In some plots they are almost indistinguishable from the nominal value (solid yellow line).}
    \label{fig:cornerplots_validation}
\end{figure*}

\begin{figure*}[h]
    \centering

    \begin{subfigure}[b]{0.42\textwidth}
        \centering
        \includegraphics[width=\linewidth,height=\linewidth]{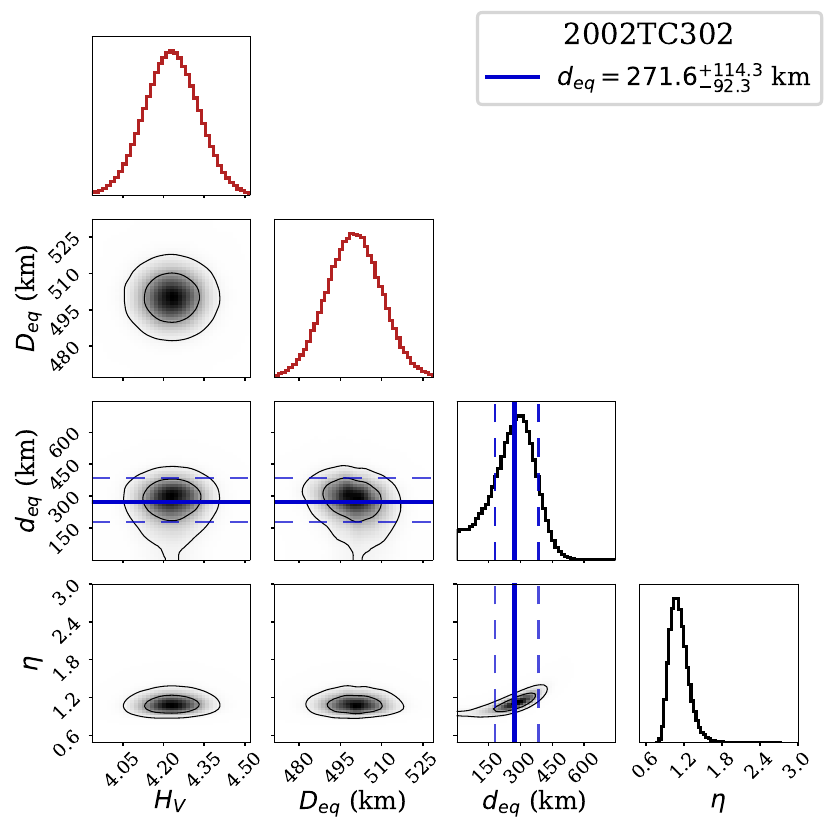}
    \end{subfigure}
    \hspace{1.5cm}
    \begin{subfigure}[b]{0.42\textwidth}
        \centering
        \includegraphics[width=\linewidth,height=\linewidth]{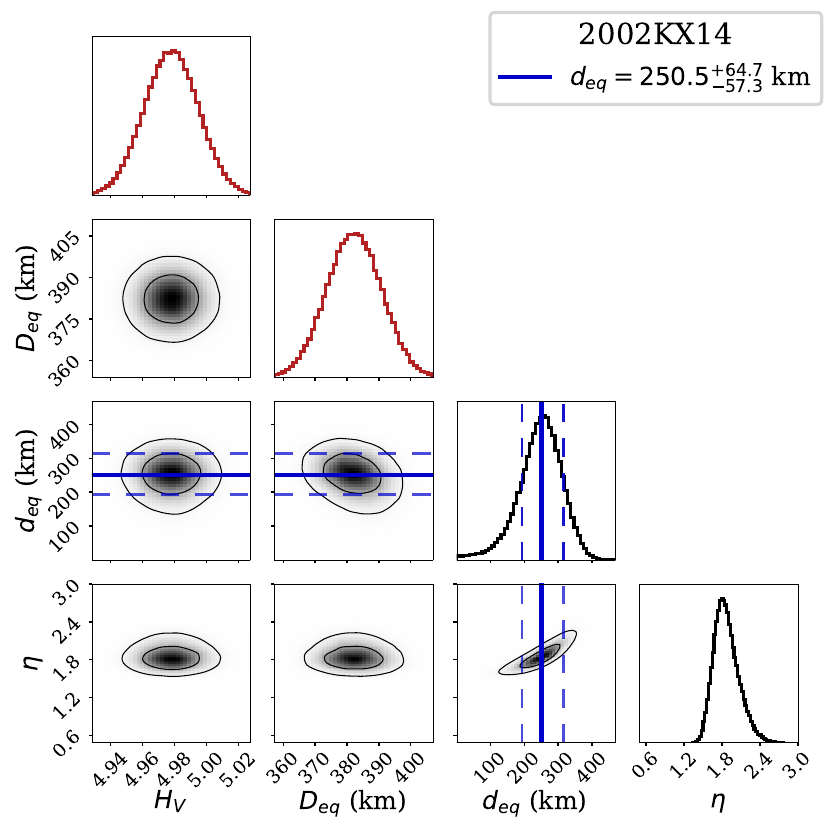}
    \end{subfigure}

    \begin{subfigure}[b]{0.42\textwidth}
        \centering
        \includegraphics[width=\linewidth,height=\linewidth]{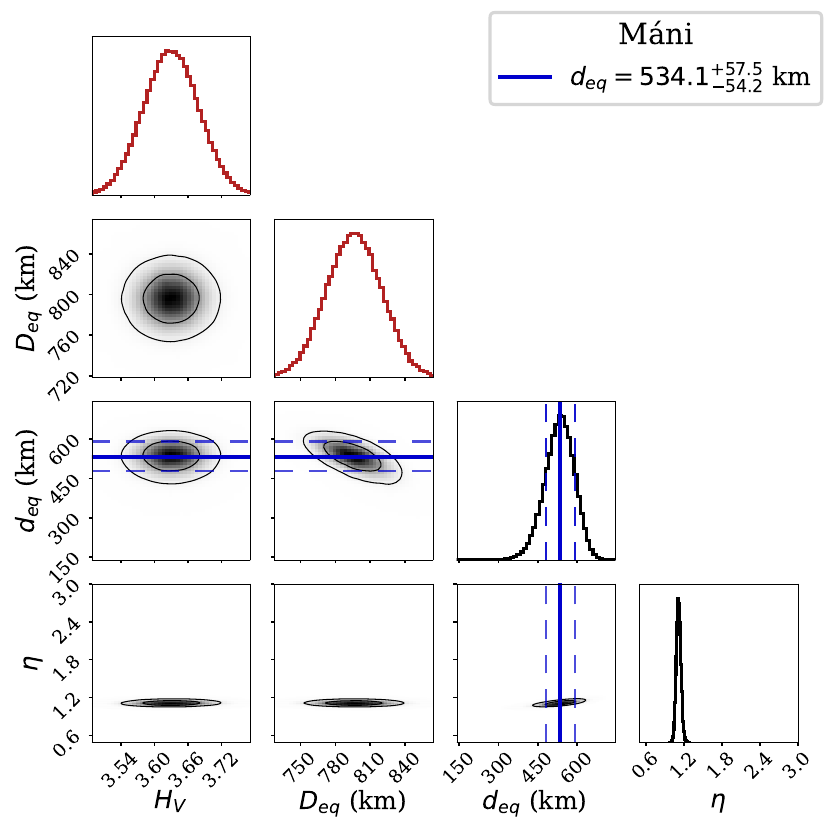}
    \end{subfigure}
    \hspace{1.5cm}
    \begin{subfigure}[b]{0.42\textwidth}
        \includegraphics[width=\linewidth,height=\linewidth]{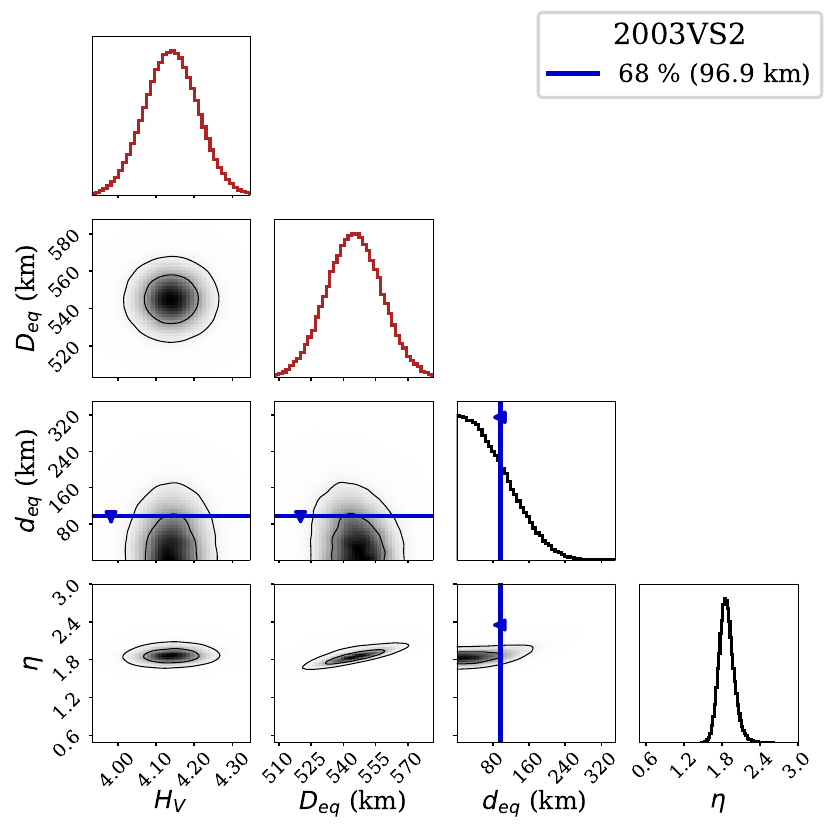}
    \end{subfigure}

    \begin{subfigure}[b]{0.42\textwidth}
        \centering
        \includegraphics[width=\linewidth,height=\linewidth]{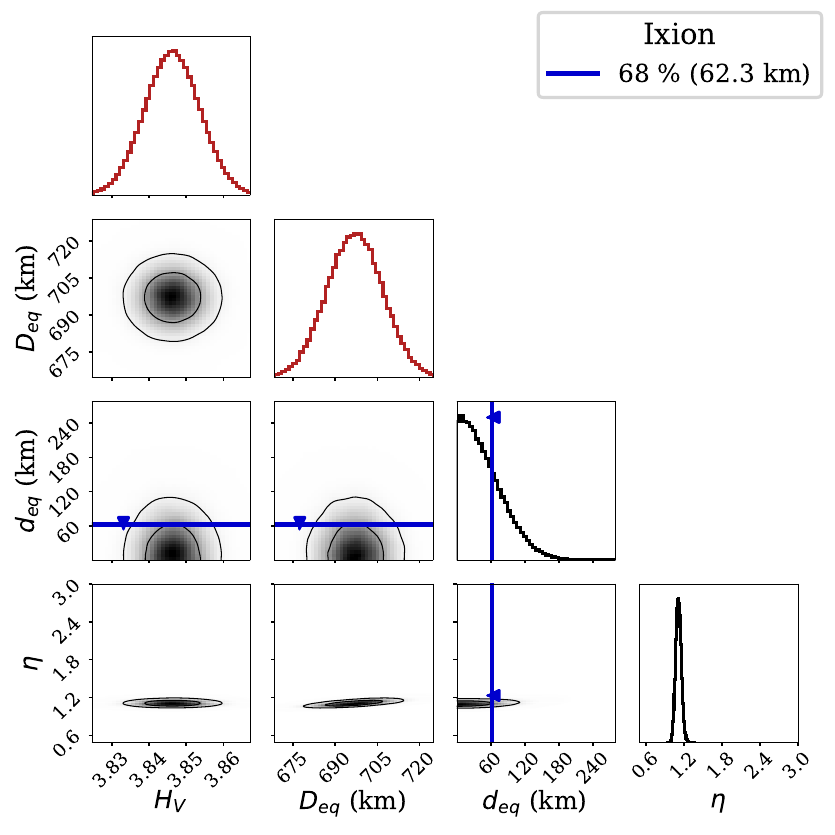}
    \end{subfigure}
    \hspace{1.5cm}
    \begin{subfigure}[b]{0.42\textwidth}
        \centering
        \includegraphics[width=\linewidth,height=\linewidth]{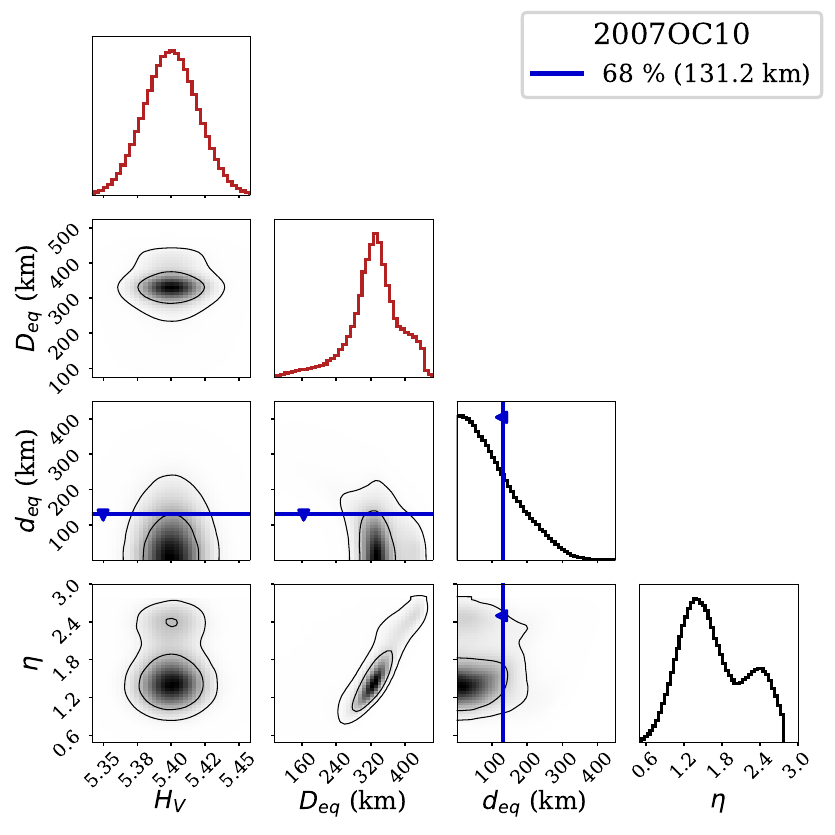}
    \end{subfigure}
    \caption{Pairwise plots for non-validation targets. The marginalised posterior pdfs of $H_V$ and $D_{\text{eq}}$ are shown in red to highlight the fact that these are not free parameters. These pdfs represent the independent measurements of $H_V$ and $D_{\text{eq}}$ from literature, reported in Table \ref{tab:merged_table}, that were included in the analysis.}
    \label{fig:cornerplots_rest}
\end{figure*}

 Our results are summarised in Table \ref{tab:results_summary}. In Figure \ref{fig:summary} we graphically represent the obtained satellite sizes. The  pairwise plots for the obtained posterior samples are shown in Figure \ref{fig:cornerplots_validation} for the validation targets and in Figure \ref{fig:cornerplots_rest} for the remaining targets. Plots showing the measured thermal fluxes against the modelled thermal spectral energy distribution are presented in Appendix \ref{ap:results}.

\subsection{\achlyslong{}}

We infer the presence of a companion for \achlys{}. This is consistent with the fact that \achlys{} has a known satellite, see Section \ref{sec:sampleselection}. However, the estimated satellite diameter of $77^{+10}_{-11}$~km from the observed magnitude difference \citep{dias2017study} is smaller than the value obtained with our methodology $d_{\text{eq}}=236^{+64}_{-52}$~km. It is important to note that we do not impose a difference in brightness between the primary and the secondary in our methodology. HST measured it to be of $\Delta m = 5.0 \pm 0.3$ magnitudes, whereas our results imply $\Delta m = 2.6 ^{+0.7}_{-0.5}$ mag. We hypothesise three scenarios to solve this discrepancy. The first one is that there is a considerable difference in albedos and/or beaming parameters between \achlys{}' main body and the satellite, whereas in our modelling we consider equal surface properties for both objects. The second one is that \achlys{}' rotationally averaged area-equivalent diameter has been underestimated by $5$\%. This would propagate to a nominally inferred satellite size of $85$~km, given the negative correlation between $D_{\text{eq}}$ and $d_{\text{eq}}$ observed in our posterior sample. Finally, it might be the case that there is a second previously undetected satellite in the system. In our modelling, to account for the extra flux from this putative second satellite we need a larger value of $d_{\text{eq}}$.

\subsection{\gkunlong{}}

For \gkun{} our analysis yields a marginalized posterior in $d_{\text{eq}}$ peaking at 0. We rule out the presence of any satellite above 159.1 ~km in diameter with 68\% credibility, and above 258.8 ~km in diameter with 90\% credibility. G!\`{o}'\'{e}!h\'{u}, the known satellite of \gkun{}, is estimated to be $139.5 \pm 2.9$ ~km in diameter from the magnitude difference reported in \cite{grundy2019Gkunhondima} and the observed size of \gkun from occultations (assuming equal albedos), which is consistent with our derived 68\% credibility upper limit.

This is the only target with a known satellite for which we do not recover its presence. The scarce thermal wavelength coverage, with only \textit{Herschel}/PACS data, leads to a wide $d_{\text{eq}}$ marginalised posterior pdf, preventing us from setting tighter constraints.

\subsection{\huyalong{}{}}

For \huya{} we infer the presence of a satellite $217^{+25}_{-22}$ ~km in diameter. The marginalised posterior in $d_{\text{eq}}$ shows a negligible posterior probability of the satellite being smaller than 100 ~km in diameter (see Figure \ref{fig:cornerplots_validation}). This is the TNO in our sample with smallest uncertainties in $d_{\text{eq}}$ for those falling in case 2 of the distinction made in Section \ref{sec:Analysis}. The size of the satellite obtained in this work is in agreement with the lower limit of 165 ~km reported in \cite{rommel2025stellar}, from direct measurements with stellar occultations.

\subsection{\vardalong{}{}}

With \varda{}, we predict the presence of a satellite of size $d_{\text{eq}}=332^{+181}_{-158}$ ~km. The $1\sigma$ credible interval in $d_{\text{eq}}$ spans more than 300 ~km, and the corresponding marginalised posterior does not drop to 0 approaching $d_{\text{eq}}=0$, see Figure \ref{fig:cornerplots_validation}. The results are less conclusive in this case than for other objects, since the thermal data wavelength coverage is narrower, lacking \textit{Spitzer}/MIPS and \textit{Herschel}/SPIRE measurements. In spite of this, the result is compatible with the estimated size of the known satellite Ilmarë, of $332.9^{+9.1}_{-8.8}$ ~km in diameter from the difference in magnitudes reported in \cite{grundy_mutual_2015}.

\subsection{\tclong{}}

The occultation by \tc{} revealed an area-equivalent diameter $\sim 84$ ~km smaller than that inferred by the fitting of a single-object NEATM model to the thermal data \citep{ortiz2020large, fornasier_tnos_2013}. It has been suggested that this difference could be attributed to an undetected satellite. By simply equating $D^2_{\text{eq, occ}} + d^2_{\text{eq}}  = D^2_{\text{eq, thermal}}$, a rough estimate of the size of 100-300 ~km was given in \cite{ortiz2020large} for this putative satellite. From our more rigorous analysis, we obtain a $1\sigma$ credible interval for $d_{\text{eq}} $ of [179, 386] ~km, reasonably compatible with this previous estimate.

In \cite{ortiz2020large}, the effects of a putative satellite in this size range on the short term photometry and astrometric residuals are explored. They conclude that a $\sim 200$ ~km secondary could account for the dispersion observed in the astrometric residuals and the observed rotational photometric variability. In the same work, the authors report that the observational coverage in the cross-track direction away from the main body was poor. Therefore, a satellite in the proposed size range could have easily been missed.

It is worth noting that the posterior pdf is non-zero for $d_{\text{eq}} = 0$, see Figure \ref{fig:cornerplots_rest}. This indicates that our analysis is not completely incompatible with \tc{} being a single object, although the presence of a sizeable satellite is more likely.

\subsection{\kxlong{}}

Similarly to the case of \tc{}, the area-equivalent diameter from occultation data is found to be $\sim 65$ ~km smaller than that from single-object model fitted to thermal measurements \citep{rizos2025trans}. In our analysis, the presence of a sizeable satellite is inferred, with a credible interval of $1\sigma$ for its diameter of [194, 316] ~km. In this case, the marginal posterior in $d_{\text{eq}}$ is almost zero for small $d_{\text{eq}}$ values (see Figure \ref{fig:cornerplots_rest}), which means that evidence for the presence of a putative satellite is strong according to our methodology. Our modelling of this TNO shows a remarkably good fit to the thermal data, see Figure \ref{fig:emission_group2}.

Four different occultations have been observed for this object, and none show secondary detections compatible with such a satellite. Nevertheless, the positive chords always scan \kx{}'s sky plane in almost the same direction (see Fig. B.6 in \cite{rizos2025trans}), and there are in total only two negative observations constraining the surroundings of the body.

\subsection{\manilong{}{}}

This object shows the largest deviation between the diameter estimated from thermal data (single-object modelling) and the diameter directly measured from occultations, being the latter $\sim 138$ ~km smaller than the former \citep{rommel2023large}. According to our analysis, it is also the object that should have the largest companion; the inferred $1\sigma$ credible interval is [480, 592] ~km, roughly two thirds of the diameter of the main body. Our posterior estimations strongly suggest that a large satellite should be present, as \posterior is almost null for $d_{\text{eq}} < 300$ ~km, see Figure \ref{fig:cornerplots_rest}.

\mani{} has unprecedented occultation data coverage. Nine occultations are published in \cite{rommel2023large}, where three are single-chord, four are double-chord and two are multi-chord, with three and 61 positive detections respectively. However, no secondary features compatible with a satellite are detected. The surroundings of the main body are explored with great resolution in the event with 61 positives (8 August 2022), as many negative light curves are available near the main detection. These negatives scan an area 900 ~km to the South and 1100 ~km to the North from the centre of the detected (main) body, see \cite[Figure 3]{rommel2023large}. Similarly, the along-track coverage also covers more than 1000 ~km East and West of the main body. The chords show spacings of, at most, $\sim 200$ ~km between them.

\subsection{\vslong{}}

For \vs, we obtain a marginalised posterior in $d_{\text{eq}}$ peaking at 0, see Figure \ref{fig:cornerplots_rest}. Therefore, we derive an upper limit on the presence of a putative companion. We rule out the presence of any satellite above 97~km in diameter with 68\% credibility, and above 155~km in diameter with 90\% credibility.
The rotationally averaged occultation based diameter $D_\text{eq}=545\pm 24$ ~km and derived beaming parameter $\eta = 1.87 ^{+0.10}_{-0.12}$ are consistent with the values of the thermal analysis reported in \cite{mommert_tnos_2012} of $D_\text{eq}=523.0^{+35.1}_{-34.4}$ ~km, $\eta = 1.57^{+0.30}_{-0.23}$ within the uncertainties.

\subsection{\ixionlong{}{}}

For \ixion{}, no satellite is needed to reconcile the occultation and thermal measurements, since the marginalised posterior in $d_{\text{eq}}$ peaks at 0. \cite{lellouch_tnos_2013} report a fit of the single body NEATM model with $D_{\text{eq}}=617^{+19}_{-20}$ ~km and $\eta = 0.91^{+0.04}_{-0.06}$. The area-equivalent diameter measured with occultations of $D_{\text{eq}}=697\pm 10$ ~km is considerably larger. Therefore, we infer a beaming parameter of $\eta=1.11^{+0.04}_{-0.06}$, larger than \cite{lellouch_tnos_2013}, as an increasing beaming parameter implies a colder temperature distribution (see Eqs. \ref{eq:temperature_distribution} and \ref{eq:Tss}), and consequently diminished thermal emission.

\subsection{\oclong{}}

For \oc{} we find a 90\% credibility upper limit on the area-equivalent diameter for a putative satellite of 218 ~km. As was the case for \varda{}, the limited wavelength coverage of the thermal emission combined with a large $\sim 15$\% relative errorbar in the occultation derived value of $D_{\text{eq}}$, prevents us from setting tighter constraints.

Our results are almost identical to those of \cite{gomez2025size} (see their Figure 6), where the reported  90\% credibility upper limit on $d_{\text{eq}}$ is 225.3 ~km. The only difference in approaches is that we incorporate the uncertainties in $H_V$, whereas in \cite{gomez2025size} the absolute magnitude is fixed to the nominal value of $H_V = 5.4$.

As was the case for \vs, the occultation based diameter $D_{\text{eq}} = 330^{+56}_{-55}$ ~km and our estimated beaming parameter of $\eta=1.59^{+0.48}_{-0.61}$ are compatible within $1\sigma$ those fitted in \cite{lellouch_tnos_2013} with a single-object NEATM: $D_{\text{eq}}=306^{+93}_{-72}$ ~km, $\eta=1.17^{+1.11}_{-0.64}$.

\begin{table*}

\caption{Summary of our results.}
\label{tab:results_summary}
\renewcommand{\arraystretch}{1.3} 
\centering
\begin{tabular}{lp{1.3 cm}p{1.3 cm}p{1.7 cm}p{1.6 cm}cc}
\hline \hline
Object & Satellite inferred & Known satellite &  Observed $d_{\text{eq}}$ (km) & Inferred $d_{\text{eq}}$ (km) &  Inferred $p_V$ & Inferred $\eta$ \\
\midrule
\achlyslong{}{} & Yes & Yes & $77^{+10}_{-11}$ & $236^{+64}_{-52}$ & $0.083^{+0.009}_{-0.009}$ & $1.18^{+0.04}_{-0.04}$ \\
\gkunlong{}{} & No & Yes & $139.5^{+2.9}_{-2.9}$ & <159.1& $0.146^{+0.017}_{-0.017}$ & $1.43^{+0.13}_{-0.18}$ \\
\huyalong{}{} & Yes & Yes & $>165^{*}$ & $217^{+25}_{-22}$  & $0.079^{+0.004}_{-0.004}$  & $0.98^{+0.03}_{-0.03}$\\
\vardalong{}{} & Yes & Yes & $332.9^{+9.1}_{-8.8}$ & $332^{+184}_{-158}$  & $0.097^{+0.017}_{-0.014}$ & $0.91^{+0.14}_{-0.20}$ \\
\midrule
\tclong{}{}  & Yes &No&- & $272^{+114}_{-92}$ & $0.111^{+0.019}_{-0.022}$ & $1.11^{+0.12}_{-0.16}$\\
\kxlong{}{} & Yes & No&- & $251^{+65}_{-57}$ & $0.084^{+0.011}_{-0.013}$ & $1.85^{+0.16}_{-0.21}$ \\
\manilong{}{}   & Yes & No&- &   $534^{+58}_{-54}$ & $0.068^{+0.004}_{-0.005}$ & $1.12^{+0.03}_{-0.04}$ \\
\vslong{}  & No & No&- & <96.9 &$0.128^{+0.010}_{-0.011}$ & $1.87 ^{+0.10}_{-0.12}$\\
\ixionlong{}{}  & No & No &- & <62.3  & $0.104^{+0.003}_{-0.003}$ & $1.11^{+0.04}_{-0.05}$ \\
\oclong{}{} & No & No &- & <131.2  & $0.100^{+0.017}_{-0.036}$ & $1.59^{+0.48}_{-0.61}$ \\
\bottomrule
\end{tabular}
\tablefoot{A target is marked as "Satellite inferred" if the posterior in $d_{\text{eq}}$ peaks at a non-zero value. For details see Section \ref{sec:Analysis}. When an upper limit is reported, it corresponds to 68 \% credibility. For our validation targets (except Huya), the reported literature $d_{\text{eq}}$ value is computed from the observed magnitude difference (see Section \ref{sec:sampleselection}) and the measured $D_{\text{eq}}$ from occultations (see Table \ref{tab:merged_table}), under the assumption of equal albedos for both components. ${}^*$ For Huya, the literature lower limit comes from a direct detection from a stellar occultation \citep{rommel2025stellar}.}
\end{table*}

\subsection{General results}
\label{sec:General_results}

Our analysis confirms an expected negative correlation between $d_{\text{eq}}$ and $D_{\text{eq}}$, physically representing the trade-off required to maintain a constant integrated thermal flux. For objects where a satellite is inferred, ordinary least squares regression yields steep slopes between $-1.5$ and $-2.0$ (with \achlys{} reaching a marked $-3.9$). In contrast, targets compatible with a single-body model exhibit a much weaker dependence, with slopes ranging only from $-0.1$ to $-0.7$. This inverse relationship dictates that as the primary’s size decreases, the secondary’s size must increase to account for the total observed flux.

Because of this correlation, results for objects with large rotational variability (leading to uncertain $D_{\text{eq}}$ values) should be taken with caution, especially $d_{\text{eq}}$ estimates when a satellite is inferred. However, for our sample of TNOs we have small amplitudes for their rotational light curves, see Table \ref{tab:merged_table}. This is an indication that the projected area should not vary much along rotation. For reference, assuming that the reflected flux is directly proportional to the projected area, a light curve amplitude of 0.1 mag translates to $\sim $ 5\% rotational variability in projected area-equivalent diameter. 

The only object with a notably high rotational light curve amplitude is \vs. For this TNO we have used as input value for $D_{\text{eq}}$ the mean equivalent diameter estimated in \cite{vara2022multichord} by combining the projected area in the occultation with the known rotational phase at the occultation instant. This choice should mitigate the impact that rotational effects might cause on our inference on the value of $d_{\text{eq}}$.

From the pairwise plots available in Figures \ref{fig:cornerplots_validation} and \ref{fig:cornerplots_rest}, we can see that the free parameter $d_{\text{eq}}$  and the absolute magnitude are uncorrelated. This is reflected in the fact that the results presented here for \oc{} are virtually the same as those exposed in \cite{gomez2025size}, where uncertainties in $H_V$ were not considered.

\section{Conclusions}\label{sec:conclusions}

In this paper, we have developed and implemented a Bayesian framework to constrain the presence of unresolved satellites around Trans-Neptunian Objects, independent of primary-to-secondary separation. By combining radiometric thermal data with high-precision size measurements from stellar occultations, we provide a systematic approach that advances the field beyond previous qualitative or non-uniform analyses. Our methodology exploits the discrepancy between the area-equivalent diameters derived from occultations and the larger effective diameters often inferred from single-object thermal models. By rigorously modelling the observed thermal fluxes with a binary object model, we are able to indirectly detect companions and constrain their sizes in a way that is both reproducible and scalable to larger TNO populations. The main conclusions of this study can be summarised as follows:

\begin{itemize}
    \item For three out of the four targets with known satellites, we recover the presence of a companion. For the case of the target with the smallest satellite, G!kún\textdoublepipe\textprimstress hòmdímà we report an upper limit for $d_{\text{eq}}$ consistent with the estimated size of its satellite  G!\`{o}'\'{e}!h\'{u}. Therefore, our methodology delivers coherent results for these validation targets.
    
    In the case of \achlys{}, the reported satellite size is larger than expected from the measured difference in brightness between primary and secondary. This could be due to a  $\sim 5$\% underestimation  of $D_{\text{eq}}$ from occultation data. The results could also possibly be reconciled if this difference in brightness constraint is imposed and an extra degree of freedom is added to the model, allowing the primary and secondary to have different albedos. Another possibility is to model \achlys{} as a triple system. However, this is beyond the scope of the present paper.
    
    For \huya{} and \varda{}, our estimated satellite sizes are compatible with the estimates available in the literature. This is especially encouraging for the case of \huya{}, which has complete thermal emission wavelength coverage within "TNOs are Cool", a very accurate measurement of $D_{\text{eq}}$, and a reliable constraint on the real value of $d_{\text{eq}}$ from stellar occultations.

    \item We find that for four objects (including  G!kún\textdoublepipe\textprimstress hòmdímà), no satellite is needed to reconcile thermal and occultation observations, and we provide upper limits on the size of any putative companion. For the other six objects, we find that the presence of a satellite provides a better explanation of the measured fluxes from "TNOs are Cool", taking into account the observed occultation sizes.

    This corresponds to a nominal binarity fraction of 60\% within our sample. However, we emphasize that this value is derived from a small ($N = 10$) and heterogeneous set of objects, and is therefore not directly comparable to population-level estimates. For reference, \cite{porter_detection_2024} report a binary fraction of 21.2\% based on a much larger ($N = 198$) and homogeneous sample of Cold Classical TNOs observed with HST.

    \item Our analysis provides compelling evidence that \manilong{}, \tclong{}, and \kxlong{} are TNBs. For these targets, the size constraints from multi-chord occultations are very reliable thanks to great observational coverage and low expected rotational variability. We infer significant sizes for the putative satellites of these targets, with primary-to-secondary diameter ratios ranging from 3:2 to 2:1. These ratios represent an intermediate regime between the near-equal-sized binaries typically produced via Streaming Instability models and the high-mass-ratio systems composed of large TNOs with smaller moons. However, these size ratio estimates must be taken with care as they come from our assumption of equal surface properties.
    
    These objects have been targeted by both HST imaging and stellar occultations, but no companions have been reported to date. However, several geometric configurations and small angular separations could explain why such satellites remained undetected in previous observations. Further observations are essential to confirm these findings. Stellar occultations with broad spatial coverage represent the most promising technique for this purpose.

    \item In Appendix \ref{ap:robustness}, we quantitatively assessed the impact of non-spherical geometries and different emissivity values on our results. Our analysis demonstrates that the observed excess thermal flux in the targets for which we infer the presence of a satellite cannot be explained with values of the emissivity  different from $\epsilon=0.9$. Crucially, we also prove that the spherical  approximation within the NEATM does not introduce a systematic bias toward smaller fluxes, leading to an overestimation of satellite sizes. Instead, plausible ellipsoidal models yield a scatter of flux values around the spherical prediction. While this increases the variance, it does not result in a shift that would lead to systematic false-positive satellite detections. We acknowledge that for \kx{} and \tc{}, certain triaxial geometries could potentially account for the observed thermal emission. However, fully constraining a triaxial shape-spin model involves a significantly larger parameter space and a level of complexity that remains outside the scope of this work. At present, the available occultation data for these targets are insufficient to faithfully constrain the numerous degrees of freedom of such a triaxial spin-shape model. Nevertheless, our current statistical framework suggests that, under the assumption of random pole orientations, the observed flux excesses are likely not due to departures from sphericity.

    Rotational variability could be an influential factor for targets with shapes departing considerably from sphericity. For the cases of \achlys{} and \vs{}, we addressed this issue by using rotationally averaged $D_{\text{eq}}$ values. For the rest of targets, rotational variability is expected to be very low from their light curve amplitudes.

    \item The thermal wavelength coverage is very influential in the accuracy of the results from our methodology. The uncertainties obtained for targets with only \textit{Herschel}/PACS data are significantly larger than for those with also \textit{Spitzer}/MIPS and \textit{Herschel}/SPIRE data. \varda{} and \huya{} represent a clear example: Both objects show $\sim 2$\% relative uncertainty in the occultation-derived diameter (see Table \ref{tab:merged_table}). For \varda{}, whose thermal emission was only covered by \textit{Herschel}/PACS, the obtained relative uncertainty in $d_{\text{eq}}$ is $\sim 50$\% and in $\eta$ it is $\sim 20$\%. Conversely, \huya{} was also observed with \textit{Spitzer}/MIPS,  \textit{Herschel}/SPIRE and ALMA, and shows relative uncertainties of $\sim 10$\% and $\sim 3$\% in $d_{\text{eq}}$ and $\eta$ respectively.

    \item Recent work by \cite{lyra_where_2025} highlighted a "dearth" of trans-Neptunian binaries between absolute magnitudes 4 and 5. While many individual TNOs occupy this range, the lack of known binaries suggests either primordial disruption or an observational bias toward undetected companions. Distinguishing between these scenarios is critical. If the gap is physical, \cite{lyra_where_2025} propose a hypothesis where pebble accretion stripped primordial companions for the planetesimals closest to the sun, while driving their growth. These objects will result in the current high-mass TNBs, whose satellites should be products of recent collisions. In contrast, smaller objects that did not undergo significant pebble accretion resulted in the currently known smaller TNBs, which are primordial.
    
    The low mass-limit edge of the gap seems to be consistent with the high-end mass of the  size distribution of Cold Classical TNOs \citep{kavelaars2021ossos}, a dynamical class where many equal-sized binaries are present. The high-mass end is of uncertain origin, with binaries above this edge being mainly hot classicals, scattered disk or resonant objects with smaller satellites in tight orbits. There are only two currently known TNBs in the $4 < H < 5$ range: (120347) Salacia (scattered disk object) and (82075) 2000 YW134 (resonant object). They share characteristics with the population of larger TNBs, being the primary-to-secondary diameter ratio roughly 3:1 for both and a separation on the order of $10^{-3}$ Hill radii (see \cite{johnston2019binary} and references therein).

    With our methodology, we find that \tclong{} and \kxlong{} are likely binary systems, and they are both in the 4th to 5th absolute magnitude range, see Figure \ref{fig:summary}. \tc{} is a resonant TNO, whereas \kx{} is a classical TNO with very low inclination \citep{rizos2025trans}, that could be classified as cold classical although it might also be part of the low-inclination tail of the hot population \citep{fernandez2021compositional}. The inferred size ratios and potentially tight orbit (consistent with no-detection from HST imaging) indicate that these systems also follow the trend of TNBs above the high-mass end of the gap.
    
    Confirmation of these putative satellites would double the known TNB population within this magnitude range, suggesting the gap is an artifact of observational bias rather than physical disruption. This would challenge the hypothesis that massive TNBs with small companions are exclusively products of recent collisions, favouring instead a scenario where TNOs retain primitive satellites which are below current detection thresholds.

    \item Beyond the shape constraints previously discussed, the NEATM framework may be inherently limited for certain targets due to unmodeled physical processes. The presence of undetected ring systems, for instance, could contribute significant thermal emission that is not captured by standard single or binary models \citep{lellouch2017thermal, muller2019haumea, kiss_visible_2024}. Furthermore, while less probable, the thermal energy balance of some objects could be perturbed by the sublimation of surface ices or active cryovolcanism \citep{kiss_prominent_2024}. The fact that neither the single-body models in existing literature nor the binary configurations proposed in this work perfectly reproduce the full multi-wavelength spectral energy distribution for every target suggests that simplified thermal approximations have reached their predictive limit. These remaining discrepancies likely reflect a combination of complex surface physics and the inherent systematic uncertainties associated with the reduction and calibration of low-signal-to-noise thermal data.

    \item Finally, this work highlights the critical importance of obtaining multi-chord stellar occultations for targets of the "TNOs are Cool" survey, especially if they have broad thermal wavelength coverage. The synergy between occultation-derived diameters and radiometric data significantly enhances the scientific value of the legacy measurements provided by \textit{Herschel} and \textit{Spitzer}. It is important to emphasise that sensitive photometric observations at $\sim$100 \micron{}, where the thermal emission of TNOs typically peaks, will remain unavailable for the near future, as no current space-borne or ground-based facilities cover this spectral range. Consequently, it is essential to maximise the scientific return of the existing "TNOs are Cool" dataset with approaches such as the one presented here.

\end{itemize}

\begin{acknowledgements}
The authors thank the referee and the editor for their valuable feedback. This publication is funded by the Spanish Ministry of Universities through the university training programme FPU2022/00492. We acknowledge Flavia Luane Rommel for her insights on \mani{}. We acknowledge the use of the publicly available Python wrapper for NEATM modelling developed by Michael Müller. The authors of IAA-CSIC acknowledge financial support from the Severo Ochoa grant CEX2021-001131-S funded by MCIN/AEI/ 10.13039/501100011033. This work was partially funded by the Spanish projects PID2020-112789GBI00 (AEI) and Proyecto de Excelencia de la Junta de Andalucía PY20-01309. P.S-S. and Y.K. acknowledge financial support from the Spanish I+D+i project PID2022-139555NB-I00 (TNO-JWST) funded by MCIN/AEI/10.13039/501100011033. This research has made use of LTE's SsODNet VO service (https://ssp.imcce.fr/webservices/ssodnet/).
\end{acknowledgements}


\bibliographystyle{aa}
\bibliography{references}

@article{nesvorny2018dynamical,
  title={Dynamical evolution of the early solar system},
  author={Nesvorný, David},
  journal={Annual Review of Astronomy and Astrophysics},
  volume={56},
  number={1},
  pages={137--174},
  year={2018},
  publisher={Annual Reviews}
}

@article{christy1978satellite,
  title={The satellite of Pluto},
  author={Christy, James W and Harrington, Robert Sutton},
  journal={Astronomical Journal, vol. 83, Aug. 1978, p. 1005, 1007, 1008.},
  volume={83},
  pages={1005},
  year={1978}
}

@article{leinhardt2010formation,
  title={The formation of the collisional family around the dwarf planet Haumea},
  author={Leinhardt, Zo{\"e} M and Marcus, Robert A and Stewart, Sarah T},
  journal={The Astrophysical Journal},
  volume={714},
  number={2},
  pages={1789},
  year={2010}
}

@article{parker2016discovery,
  title={Discovery of a Makemakean moon},
  author={Parker, Alex H and Buie, Marc W and Grundy, Will M and Noll, Keith S},
  journal={The Astrophysical Journal Letters},
  volume={825},
  number={1},
  pages={L9},
  year={2016}
}

@article{gomez2025size,
  title={Size and shape of the trans-Neptunian object (470316) 2007 OC10: Comparison with thermal data},
  author={G{\'o}mez-Lim{\'o}n, Jos{\'e} Mar{\'\i}a and Leiva, Rodrigo and Ortiz, JL and Morales, Nicol{\'a}s and Kretlow, Mike and Vara-Lubiano, M{\'o}nica and Santos-Sanz, Pablo and Alvarez-Candal, Alvaro and Rizos, Juan L and Duffard, Ren{\'e} and others},
  journal={Astronomy \& Astrophysics},
  volume={697},
  pages={A157},
  year={2025}
}

@article{grundy2019Gkunhondima,
  title={The mutual orbit, mass, and density of transneptunian binary Gǃk{\'u}nǁ'h{\`o}md{\'\i}m{\`a} (229762 2007 UK126)},
  author={Grundy, W. M. and Noll, K. S. and Buie, M. W. and Benecchi, S. D. and Ragozzine, D. and Roe, H. G.},
  journal={Icarus},
  volume={334},
  pages={30--38},
  year={2019},
  sortyear={2019}
}

@phdthesis{thirouin_thesis_2013,
	type = {doctoral thesis},
	title = {Study of transneptunian objects using photometrics techniques and numerical simulations},
	copyright = {Creative Commons Attribution-NonCommercial-NoDerivs 3.0 License},
	language = {eng},
	urldate = {2025-12-22},
	school = {Universidad de Granada},
	author = {Thirouin, Audrey},
	year = {2013}
}

@article{grundy_mutual_2015,
	title = {The mutual orbit, mass, and density of the large transneptunian binary system {Varda} and {Ilmarë}},
	volume = {257},
	issn = {0019-1035},
	journal = {Icarus},
	author = {Grundy, W. M. and Porter, S. B. and Benecchi, S. D. and Roe, H. G. and Noll, K. S. and Trujillo, C. A. and Thirouin, A. and Stansberry, J. A. and Barker, E. and Levison, H. F.},
	year = {2015},
	pages = {130--138},
}

@article{noll201238628,
  title={(38628) Huya},
  author={Noll, KS and Grundy, WM and Schlichting, H and Murray-Clay, R and Benecchi, SD},
  journal={International Astronomical Union Circular},
  volume={9253},
  pages={2},
  year={2012}
}

@article{sheppard2007light,
  title={Light curves of dwarf plutonian planets and other large Kuiper belt objects: Their rotations, phase functions, and absolute magnitudes},
  author={Sheppard, Scott S},
  journal={The Astronomical Journal},
  volume={134},
  number={2},
  pages={787},
  year={2007}
}

@article{benedetti2016results,
  title={Results from the 2014 November 15th multi-chord stellar occultation by the TNO (229762) 2007 UK126},
  author={Benedetti-Rossi, Gustavo and Sicardy, Bruno and Buie, Marc W and Ortiz, Jose L and Vieira-Martins, Roberto and Keller, Jennifer M and Braga-Ribas, Felipe and Camargo, Julio IB and Assafin, Marcelo and Morales, Nicolas and others},
  journal={The Astronomical Journal},
  volume={152},
  number={6},
  pages={156},
  year={2016}
}

@article{kiss2014optimized,
  title={Optimized Herschel/PACS photometer observing and data reduction strategies for moving solar system targets},
  author={Kiss, Cs and M{\"u}ller, Th G and Vilenius, E and P{\'a}l, A and Santos-Sanz, P and Lellouch, E and Marton, G and Vereb{\'e}lyi, E and Szalai, N and Hartogh, P and others},
  journal={Experimental Astronomy},
  volume={37},
  number={2},
  pages={161--174},
  year={2014}
}

@article{perna2010colors,
  title={Colors and taxonomy of Centaurs and trans-Neptunian objects},
  author={Perna, Davide and Barucci, Maria Antonella and Fornasier, Sonia and Demeo, Francesca E and Alvarez-Candal, Alvaro and Merlin, Fr{\`e}d{\`e}ric and Dotto, Elisabetta and Doressoundiram, Alain and de Bergh, Catherine},
  journal={Astronomy \& Astrophysics},
  volume={510},
  pages={A53},
  year={2010}
}

@article{perna2013photometry,
  title={Photometry and taxonomy of trans-Neptunian objects and Centaurs in support of a Herschel key program},
  author={Perna, D and Dotto, E and Barucci, MA and Epifani, E Mazzotta and Vilenius, E and Dall’Ora, M and Fornasier, S and M{\"u}ller, TG},
  journal={Astronomy \& Astrophysics},
  volume={554},
  pages={A49},
  year={2013}
}

@article{ortiz2006short,
  title={Short-term rotational variability of eight KBOs from Sierra Nevada Observatory},
  author={Ortiz, JL and Guti{\'e}rrez, PJ and Santos-Sanz, P and Casanova, V and Sota, A},
  journal={Astronomy \& Astrophysics},
  volume={447},
  number={3},
  pages={1131--1144},
  year={2006}
}

@article{foreman-mackey_emcee_2013,
	title = {emcee : {The} {MCMC} {Hammer}},
	volume = {125},
	issn = {00046280, 15383873},
	shorttitle = {emcee},
	doi = {10.1086/670067},
	language = {en},
	number = {925},
	urldate = {2023-10-01},
	journal = {Publications of the Astronomical Society of the Pacific
},
	author = {Foreman-Mackey, Daniel and Hogg, David W. and Lang, Dustin and Goodman, Jonathan},
	month = mar,
	year = {2013},
	pages = {306--312}
}

@INCOLLECTION{belskaya2008bookchapter,
            author = {{Belskaya}, I.~N. and {Levasseur-Regourd}, A.-C. and {Shkuratov}, Y.~G. and {Muinonen}, K.},
            title = "{Surface Properties of Kuiper Belt Objects and Centaurs from Photometry and Polarimetry}",
            booktitle = {The Solar System Beyond Neptune},
            year = 2008,
            editor = {{Barucci}, M.~A. and {Boehnhardt}, H. and {Cruikshank}, D.~P. and {Morbidelli}, A. and {Dotson}, Renee},
            pages = {115-127},
            publisher = {University of Arizona Press}
}

@article{goodman_ensemble_2010,
	title = {Ensemble samplers with affine invariance},
	volume = {5},
	issn = {2157-5452},
	doi = {10.2140/camcos.2010.5.65},
	number = {1},
	urldate = {2024-06-06},
	journal = {Commun. Appl. Math. Comput. Sci.},
	author = {Goodman, Jonathan and Weare, Jonathan},
	month = jan,
	year = {2010},
	pages = {65--80}
}

@book{gregory2005bayesian,
  title={Bayesian logical data analysis for the physical sciences: A comparative approach with Mathematica{\textregistered} support},
  author={Gregory, Phil},
  year={2005},
  publisher={Cambridge University Press}
}

@article{rommel2025stellar,
  title={Stellar occultation observations of (38628) Huya and its satellite: a detailed look into the system},
  author={Rommel, FL and Fern{\'a}ndez-Valenzuela, Estela and Proudfoot, BCN and Ortiz, JL and Morgado, BE and Sicardy, Bruno and Morales, N and Braga-Ribas, F and Desmars, Josselin and Vieira-Martins, R and others},
  journal={The Planetary Science Journal},
  volume={6},
  number={2},
  pages={48},
  year={2025}
}

@article{kavelaars2021ossos,
  title={OSSOS finds an Exponential Cutoff in the Size Distribution of the Cold Classical Kuiper belt},
  author={Kavelaars, JJ and Petit, Jean-Marc and Gladman, Brett and Bannister, Michele T and Alexandersen, Mike and Chen, Ying-Tung and Gwyn, Stephen DJ and Volk, Kathryn},
  journal={The Astrophysical journal letters},
  volume={920},
  number={2},
  pages={L28},
  year={2021}
}

@article{bessell1998model,
  title={Model atmospheres broad-band colors, bolometric corrections and temperature calibrations for O-M stars},
  author={Bessell, MS and Castelli, F and Plez, Bertrand},
  journal={Astronomy \& Astrophysics},
  volume={333},
  pages={231--250},
  year={1998}
}

@article{brown2007satellites,
  title={Satellites of 2003 AZ\_84,(50000),(55637), and (90482)},
  author={Brown, ME and Suer, T-A},
  journal={International Astronomical Union Circular},
  volume={8812},
  pages={1},
  year={2007}
}

@article{fernandez2021compositional,
  title={Compositional study of trans-neptunian objects at $\lambda$> 2.2 $\mu$m},
  author={Fern{\'a}ndez-Valenzuela, Estela and Pinilla-Alonso, N and Stansberry, J and Emery, JP and Perkins, W and Van Laerhoven, C and Gladman, BJ and Fraser, W and Cruikshank, D and Lellouch, E and others},
  journal={The Planetary Science Journal},
  volume={2},
  number={1},
  pages={10},
  year={2021}
}

@article{Kilic2026Ixion,
  title={Constraining the size, shape, and albedo of the large trans-Neptunian object (28978) Ixion with multi-chord stellar occultations},
  author={Kilic, Y and Braga-Ribas, F and Pereira, CL and Ortiz, JL and Sicardy, B and Santos-Sanz, P and Erece, O and Rizos, JL and G{\'o}mez-Lim{\'o}n, JM and Margoti, G and others},
  journal={Astronomy \& Astrophysics},
  volume={707},
  pages={A70},
  year={2026}
}

@article{johnston2019binary,
  title={Binary Minor Planets Compilation V3. 0},
  author={Johnston, WR},
  journal={NASA Planetary Data System},
  pages={4},
  year={2019}
}

@article{santos2022physical,
  title={Physical properties of the trans-Neptunian object (38628) Huya from a multi-chord stellar occultation},
  author={Santos-Sanz, Pablo and Ortiz, Jos{\'e} Luis and Sicardy, Bruno and Popescu, M and Benedetti-Rossi, G and Morales, N and Vara-Lubiano, M and Camargo, JIB and Pereira, CL and Rommel, FL and others},
  journal={Astronomy \& Astrophysics},
  volume={664},
  pages={A130},
  year={2022}
}

@article{souami2020multi,
  title={A multi-chord stellar occultation by the large trans-Neptunian object (174567) Varda},
  author={Souami, D and Braga-Ribas, F and Sicardy, B and Morgado, B and Ortiz, JL and Desmars, J and Camargo, JIB and Vachier, F and Berthier, J and Carry, B and others},
  journal={Astronomy \& Astrophysics},
  volume={643},
  pages={A125},
  year={2020}
}

@article{Poglitsch2010,
  title = {The {{Photodetector Array Camera}} and {{Spectrometer}} ({{PACS}}) on the {{Herschel Space Observatory}}},
  author = {Poglitsch, A. and Waelkens, C. and Geis, N. and Feuchtgruber, H. and Vandenbussche, B. and Rodriguez, L. and Krause, O. and Renotte, E. and van Hoof, C. and Saraceno, P. and Cepa, J. and Kerschbaum, F. and Agn{\`e}se, P. and Ali, B. and Altieri, B. and Andreani, P. and Augueres, J.-L. and Balog, Z. and Barl, L. and Bauer, O. H. and Belbachir, N. and Benedettini, M. and Billot, N. and Boulade, O. and Bischof, H. and Blommaert, J. and Callut, E. and Cara, C. and Cerulli, R. and Cesarsky, D. and Contursi, A. and Creten, Y. and Meester, W. De and Doublier, V. and Doumayrou, E. and Duband, L. and Exter, K. and Genzel, R. and Gillis, J.-M. and Gr{\"o}zinger, U. and Henning, T. and Herreros, J. and Huygen, R. and Inguscio, M. and Jakob, G. and Jamar, C. and Jean, C. and de Jong, J. and Katterloher, R. and Kiss, C. and Klaas, U. and Lemke, D. and Lutz, D. and Madden, S. and Marquet, B. and Martignac, J. and Mazy, A. and Merken, P. and Montfort, F. and Morbidelli, L. and M{\"u}ller, T. and Nielbock, M. and Okumura, K. and Orfei, R. and Ottensamer, R. and Pezzuto, S. and Popesso, P. and Putzeys, J. and Regibo, S. and Reveret, V. and Royer, P. and Sauvage, M. and Schreiber, J. and Stegmaier, J. and Schmitt, D. and Schubert, J. and Sturm, E. and Thiel, M. and Tofani, G. and Vavrek, R. and Wetzstein, M. and Wieprecht, E. and Wiezorrek, E.},
  year = {2010},
  journal = {Astronomy \& Astrophysics},
  volume = {518},
  pages = {L2},
  issn = {0004-6361, 1432-0746},
  doi = {10.1051/0004-6361/201014535},
  copyright = {{\copyright} ESO, 2010}
}

@book{binzel_asteroids_1989,
	title = {Asteroids {II}},
	urldate = {2024-09-17},
	author = {Binzel, Richard P. and Gehrels, Tom and Matthews, Mildred Shapley},
	month = jan,
	year = {1989},
}

@article{muller2009tnos,
  title={TNOs are Cool: A Survey of the Transneptunian Region: A Herschel Open Time Key Programme},
  author={M{\"u}ller, Thomas G and Lellouch, Emmanuel and B{\"o}hnhardt, Hermann and Stansberry, John and Barucci, Antonella and Crovisier, Jacques and Delsanti, Audrey and Doressoundiram, Alain and Dotto, Elisabetta and Duffard, Ren{\'e} and others},
  journal={Earth, Moon, and Planets},
  volume={105},
  number={2},
  pages={209--219},
  year={2009},
  publisher={Springer}
}

@article{sicardy2011pluto,
  title={A Pluto-like radius and a high albedo for the dwarf planet Eris from an occultation},
  author={Sicardy, Bruno and Ortiz, JL and Assafin, M and Jehin, Emmanuel and Maury, A and Lellouch, Emmanuel and Hutton, R Gil and Braga-Ribas, F and Colas, Francois and Hestroffer, Daniel and others},
  journal={Nature},
  volume={478},
  number={7370},
  pages={493--496},
  year={2011}
}

@article{hovis1966infrared,
  title={Infrared reflectance spectra of igneous rocks, tuffs, and red sandstone from 0.5 to 22 $\mu$},
  author={Hovis Jr, WA and Callahan, William R},
  journal={Journal of the Optical Society of America},
  volume={56},
  number={5},
  pages={639--643},
  year={1966}
}

@article{rieke2004multiband,
  title={The multiband imaging photometer for Spitzer (MIPS)},
  author={Rieke, GH and Young, ET and Engelbracht, CW and Kelly, DM and Low, FJ and Haller, EE and Beeman, JW and Gordon, KD and Stansberry, JA and Misselt, KA and others},
  journal={The Astrophysical Journal Supplement Series},
  volume={154},
  number={1},
  pages={25--29},
  year={2004}
}

@book{gelman2013bayesian,
  title={Bayesian data analysis, Third Edition},
  author={Gelman, Andrew and Carlin, John B and Stern, Hal S and Rubin, Donald B and Vehtari, Aki and Dunson, David B},
  year={2013},
  publisher={Chapman and Hall/CRC}
}

@book{robert2007bayesian,
  title={The Bayesian choice: from decision-theoretic foundations to computational implementation},
  author={Robert, Christian P},
  year={2007},
  publisher={Springer}
}

@article{verbiscer2022diverse,
  title={The diverse shapes of dwarf planet and large KBO phase curves observed from new horizons},
  author={Verbiscer, Anne J and Helfenstein, Paul and Porter, Simon B and Benecchi, Susan D and Kavelaars, JJ and Lauer, Tod R and Peng, Jinghan and Protopapa, Silvia and Spencer, John R and Stern, S Alan and others},
  journal={The Planetary Science Journal},
  volume={3},
  number={4},
  pages={95},
  year={2022}
}

@incollection{fernandez2020introduction,
  title={Introduction: The Trans-Neptunian belt—Past, present, and future},
  author={Fern{\'a}ndez, Julio A},
  booktitle={The Trans-Neptunian Solar System},
  pages={1--22},
  year={2020},
  publisher={Elsevier}
}

@article{gladman2021transneptunian,
  title={Transneptunian space},
  author={Gladman, Brett and Volk, Kathryn},
  journal={Annual Review of Astronomy and Astrophysics},
  volume={59},
  number={1},
  pages={203--246},
  year={2021}
}

@article{thuillier2004solar,
  title={Solar irradiance reference spectra for two solar active levels},
  author={Thuillier, G and Floyd, L and Woods, TN and Cebula, R and Hilsenrath, E and Hers{\'e}, M and Labs, D},
  journal={Advances in Space Research},
  volume={34},
  number={2},
  pages={256--261},
  year={2004}
}

@article{rommel2023large,
  title={A large topographic feature on the surface of the trans-Neptunian object (307261) 2002 MS4 measured from stellar occultations},
  author={Rommel, Flavia Luane and Braga-Ribas, Felipe and Ortiz, JL and Sicardy, Bruno and Santos-Sanz, Pablo and Desmars, Josselin and Camargo, JIB and Vieira-Martins, R and Assafin, M and Morgado, BE and others},
  journal={Astronomy \& Astrophysics},
  volume={678},
  pages={A167},
  year={2023}
}

@article{schindler2017results,
  title={Results from a triple chord stellar occultation and far-infrared photometry of the trans-Neptunian object (229762) 2007 UK126},
  author={Schindler, Karsten and Wolf, J{\"u}rgen and Bardecker, Jerry and Olsen, Aart and M{\"u}ller, Thomas and Kiss, Csaba and Ortiz, Jose-Luis and Braga-Ribas, Felipe and Camargo, JIB and Herald, Dave and others},
  journal={Astronomy \& Astrophysics},
  volume={600},
  pages={A12},
  year={2017}
}

@article{vara2022multichord,
  title={The multichord stellar occultation on 2019 October 22 by the trans-Neptunian object (84922) 2003 VS2},
  author={Vara-Lubiano, M and Benedetti-Rossi, G and Santos-Sanz, P and Ortiz, JL and Sicardy, B and Popescu, M and Morales, N and Rommel, FL and Morgado, B and Pereira, CL and others},
  journal={Astronomy \& Astrophysics},
  volume={663},
  pages={A121},
  year={2022}
}

@article{benedetti2019trans,
  title={The trans-Neptunian object (84922) 2003 VS2 through stellar occultations},
  author={Benedetti-Rossi, Gustavo and Santos-Sanz, Pablo and Ortiz, Jos{\'e} Luis and Assafin, M and Sicardy, B and Morales, Nicol{\'a}s and Vieira-Martins, Roberto and Duffard, R and Braga-Ribas, F and Rommel, FL and others},
  journal={The Astronomical Journal},
  volume={158},
  number={4},
  pages={159},
  year={2019}
}

@article{dias2017study,
  title={Study of the Plutino Object (208996) 2003 AZ84 from Stellar occultations: size, shape, and topographic features},
  author={Dias-Oliveira, A and Sicardy, B and Ortiz, JL and Braga-Ribas, F and Leiva, R and Vieira-Martins, R and Benedetti-Rossi, G and Camargo, JIB and Assafin, M and Gomes-Junior, AR and others},
  journal={The Astronomical Journal},
  volume={154},
  number={1},
  pages={22},
  year={2017},
  
}

@article{rizos2025trans,
  title={The trans-Neptunian object (119951) 2002 KX14 revealed via multiple stellar occultations},
  author={Rizos, JL and Ortiz, JL and Rommel, FL and Sicardy, B and Morales, N and Santos-Sanz, P and Fern{\'a}ndez-Valenzuela, E and Desmars, J and Souami, D and Kretlow, M and others},
  journal={Astronomy \& Astrophysics},
  volume={697},
  pages={A62},
  year={2025},
  
}

@article{alvarez2016absolute,
  title={Absolute magnitudes and phase coefficients of trans-Neptunian objects},
  author={Alvarez-Candal, A and Pinilla-Alonso, N and Ortiz, Jos{\'e} Luis and Duffard, R and Morales, Nicol{\'a}s and Santos-Sanz, Pablo and Thirouin, Audrey and Silva, JS},
  journal={Astronomy \& Astrophysics},
  volume={586},
  pages={A155},
  year={2016},
  
}

@article{volk2024dynamical,
  title={Dynamical classifications of multi-opposition TNOs as of 2023 December},
  author={Volk, Kathryn and Van Laerhoven, Christa},
  journal={Research Notes of the AAS},
  volume={8},
  number={1},
  pages={36},
  year={2024}
}

@article{kretlow2024physical,
  title={Physical properties of trans-Neptunian object (143707) 2003 UY117 derived from stellar occultation and photometric observations},
  author={Kretlow, M and Ortiz, JL and Desmars, J and Morales, N and Rommel, FL and Santos-Sanz, P and Vara-Lubiano, M and Fern{\'a}ndez-Valenzuela, E and Alvarez-Candal, A and Duffard, R and others},
  journal={Astronomy \& Astrophysics},
  volume={691},
  pages={A31},
  year={2024},
  
}

@article{tegler2016two,
  title={Two color populations of Kuiper Belt and Centaur objects and the smaller orbital inclinations of red Centaur objects},
  author={Tegler, Stephen C and Romanishin, W and GJ Consolmagno, SJ},
  journal={The Astronomical Journal},
  volume={152},
  number={6},
  pages={210},
  year={2016},
  
}

@article{ortiz2020large,
  title={The large trans-Neptunian object 2002 TC302 from combined stellar occultation, photometry, and astrometry data},
  author={Ortiz, Jos{\'e} Luis and Santos-Sanz, Pablo and Sicardy, B and Benedetti-Rossi, G and Duffard, Ren{\'e} and Morales, N and Braga-Ribas, F and Fern{\'a}ndez-Valenzuela, Estela and Nascimbeni, V and Nardiello, D and others},
  journal={Astronomy \& Astrophysics},
  volume={639},
  pages={A134},
  year={2020},
  
}

@article{morgado2023dense,
  title={A dense ring of the trans-Neptunian object Quaoar outside its Roche limit},
  author={Morgado, Bruno E and Sicardy, Bruno and Braga-Ribas, F and Ortiz, JL and Salo, H and Vachier, F and Desmars, J and Pereira, CL and Santos-Sanz, P and Sfair, R and others},
  journal={Nature},
  volume={614},
  number={7947},
  pages={239--243},
  year={2023}
}

@article{ortiz2017size,
  title={The size, shape, density and ring of the dwarf planet Haumea from a stellar occultation},
  author={Ortiz, Jos{\'e} Luis and Santos-Sanz, Pablo and Sicardy, B and Benedetti-Rossi, G and B{\'e}rard, D and Morales, N and Duffard, R and Braga-Ribas, F and Hopp, U and Ries, C and others},
  journal={Nature},
  volume={550},
  number={7675},
  pages={219--223},
  year={2017},

}

@techreport{ortiz2020thermal,
  title={Thermal diameters versus occultation diameters of TNOs: a new tool to search for satellites?},
  author={Ortiz, Jose L},
  year={2020},
  institution={Copernicus Meetings}
}

@incollection{ortiz2020bookchapter,
  title={Stellar occultations by Trans-Neptunian objects: From predictions to observations and prospects for the future},
  author={Ortiz, Jos{\'e} L and Sicardy, Bruno and Camargo, Julio IB and Santos-Sanz, Pablo and Braga-Ribas, Felipe},
  booktitle={The trans-neptunian solar system},
  pages={413--437},
  year={2020},
  publisher={Elsevier}
}

@article{stansberry2012physical,
  title={Physical properties of trans-neptunian binaries (120347) Salacia--Actaea and (42355) Typhon--Echidna},
  author={Stansberry, JA and Grundy, WM and Mueller, M and Benecchi, SD and Rieke, GH and Noll, KS and Buie, MW and Levison, HF and Porter, SB and Roe, HG},
  journal={Icarus},
  volume={219},
  number={2},
  pages={676--688},
  year={2012},
  publisher={Elsevier}
}

@article{kiss2017discovery,
  title={Discovery of a satellite of the large trans-Neptunian object (225088) 2007 OR10},
  author={Kiss, Csaba and Marton, G{\'a}bor and Farkas-Tak{\'a}cs, Anik{\'o} and Stansberry, John and M{\"u}ller, Thomas and Vink{\'o}, J{\'o}zsef and Balog, Zolt{\'a}n and Ortiz, Jose-Luis and P{\'a}l, Andr{\'a}s},
  journal={The Astrophysical Journal Letters},
  volume={838},
  number={1},
  pages={L1},
  year={2017},
  
}

@article{sickafoose2019stellar,
  title={A stellar occultation by Vanth, a satellite of (90482) Orcus},
  author={Sickafoose, AA and Bosh, AS and Levine, SE and Zuluaga, CA and Genade, A and Schindler, K and Lister, TA and Person, MJ},
  journal={Icarus},
  volume={319},
  pages={657--668},
  year={2019},
  publisher={Elsevier}
}

@article{fraser2013mass,
  title={The mass, orbit, and tidal evolution of the Quaoar--Weywot system},
  author={Fraser, Wesley C and Batygin, Konstantin and Brown, Michael E and Bouchez, Antonin},
  journal={Icarus},
  volume={222},
  number={1},
  pages={357--363},
  year={2013},
  publisher={Elsevier}
}

@article{holler2021eris,
  title={The Eris/dysnomia system I: the orbit of dysnomia},
  author={Holler, Bryan J and Grundy, William M and Buie, Marc W and Noll, Keith S},
  journal={Icarus},
  volume={355},
  pages={114130},
  year={2021},
  publisher={Elsevier}
}

@article{ragozzine2009orbits,
  title={Orbits and masses of the satellites of the dwarf planet Haumea (2003 EL61)},
  author={Ragozzine, Darin and Brown, Michael E},
  journal={The Astronomical Journal},
  volume={137},
  number={6},
  pages={4766},
  year={2009},
  
}

@article{denton2025capture,
  title={Capture of an ancient Charon around Pluto},
  author={Denton, C Adeene and Asphaug, Erik and Emsenhuber, Alexandre and Melikyan, Robert},
  journal={Nature Geoscience},
  volume={18},
  number={1},
  pages={37--43},
  year={2025},

}

@article{canup2005giant,
  title={A giant impact origin of Pluto-Charon},
  author={Canup, Robin M},
  journal={Science},
  volume={307},
  number={5709},
  pages={546--550},
  year={2005},
  publisher={American Association for the Advancement of Science}
}

@article{ortiz2012rotational,
  title={Rotational fission of trans-Neptunian objects: the case of Haumea},
  author={Ortiz, JL and Thirouin, A and Campo Bagatin, A and Duffard, R and Licandro, J and Richardson, DC and Santos-Sanz, P and Morales, N and Benavidez, PG},
  journal={Monthly Notices of the Royal Astronomical Society},
  volume={419},
  number={3},
  pages={2315--2324},
  year={2012},
  publisher={The Royal Astronomical Society}
}

@article{youdin2005streaming,
  title={Streaming instabilities in protoplanetary disks},
  author={Youdin, Andrew N and Goodman, Jeremy},
  journal={The Astrophysical Journal},
  volume={620},
  number={1},
  pages={459},
  year={2005},
  
}

@article{johansen2007rapid,
  title={Rapid planetesimal formation in turbulent circumstellar disks},
  author={Johansen, Anders and Oishi, Jeffrey S and Low, Mordecai-Mark Mac and Klahr, Hubert and Henning, Thomas and Youdin, Andrew},
  journal={Nature},
  volume={448},
  number={7157},
  pages={1022--1025},
  year={2007},

}

@article{nesvorny2010formation,
  title={Formation of Kuiper belt binaries by gravitational collapse},
  author={Nesvorný, David and Youdin, Andrew N and Richardson, Derek C},
  journal={The Astronomical Journal},
  volume={140},
  number={3},
  pages={785},
  year={2010},
  
}

@article{carrera2017planetesimal,
  title={Planetesimal formation by the streaming instability in a photoevaporating disk},
  author={Carrera, Daniel and Gorti, Uma and Johansen, Anders and Davies, Melvyn B},
  journal={The Astrophysical Journal},
  volume={839},
  number={1},
  pages={16},
  year={2017},
  
}

@article{bai2010dynamics,
  title={Dynamics of solids in the midplane of protoplanetary disks: Implications for planetesimal formation},
  author={Bai, Xue-Ning and Stone, James M},
  journal={The Astrophysical Journal},
  volume={722},
  number={2},
  pages={1437},
  year={2010},
  
}

@article{thirouin2014rotational,
  title={Rotational properties of the binary and non-binary populations in the trans-Neptunian belt},
  author={Thirouin, Audrey and Noll, Keith S and Ortiz, JL and Morales, Nicolas},
  journal={Astronomy \& Astrophysics},
  volume={569},
  pages={A3},
  year={2014},
  
}

@article{muller2019haumea,
  title={Haumea’s thermal emission revisited in the light of the occultation results},
  author={M{\"u}ller, T and Kiss, Cs and Ali-Lagoa, Victor and Ortiz, Jos{\'e} Luis and Lellouch, E and Santos-Sanz, Pablo and Fornasier, S and Marton, G and Mommert, M and Farkas-Tak{\'a}cs, Anik{\'o} and others},
  journal={Icarus},
  volume={334},
  pages={39--51},
  year={2019},
  publisher={Elsevier}
}

@article{lellouch2017thermal,
  title={The thermal emission of Centaurs and trans-Neptunian objects at millimeter wavelengths from ALMA observations},
  author={Lellouch, E and Moreno, R and M{\"u}ller, T and Fornasier, S and Santos-Sanz, P and Moullet, A and Gurwell, M and Stansberry, J and Leiva, R and Sicardy, B and others},
  journal={Astronomy \& Astrophysics},
  volume={608},
  pages={A45},
  year={2017},
}

@article{brucker2009high,
  title={High albedos of low inclination Classical Kuiper belt objects},
  author={Brucker, MJ and Grundy, WM and Stansberry, JA and Spencer, JR and Sheppard, SS and Chiang, EI and Buie, MW},
  journal={Icarus},
  volume={201},
  number={1},
  pages={284--294},
  year={2009},
  publisher={Elsevier}
}

@article{veillet2002binary,
  title={The binary Kuiper-belt object 1998 WW31},
  author={Veillet, Christian and Parker, Joel Wm and Griffin, Ian and Marsden, Brian and Doressoundiram, Alain and Buie, Marc and Tholen, David J and Connelley, Michael and Holman, Matthew J},
  journal={Nature},
  volume={416},
  number={6882},
  pages={711--713},
  year={2002},

}

@article{weaver2016small,
  title={The small satellites of Pluto as observed by New Horizons},
  author={Weaver, HA and Buie, MW and Buratti, BJ and Grundy, WM and Lauer, TR and Olkin, CB and Parker, AH and Porter, SB and Showalter, MR and Spencer, JR and others},
  journal={Science},
  volume={351},
  number={6279},
  pages={aae0030},
  year={2016},
  publisher={American Association for the Advancement of Science}
}

@article{weaver2006discovery,
  title={Discovery of two new satellites of Pluto},
  author={Weaver, Harold A and Stern, SA and Mutchler, MJ and Steffl, AJ and Buie, MW and Merline, WJ and Spencer, JR and Young, EF and Young, LA},
  journal={Nature},
  volume={439},
  number={7079},
  pages={943--945},
  year={2006},

}

@article{vilenius_tnos_2012,
	title = {“{TNOs} are {Cool}”: {A} survey of the trans-{Neptunian} region: {VI}. \textit{{Herschel}} /{PACS} observations and thermal modeling of 19 classical {Kuiper} belt objects⋆},
	volume = {541},
	issn = {0004-6361, 1432-0746},
	shorttitle = {“{TNOs} are {Cool}”},
	url = {http://www.aanda.org/10.1051/0004-6361/201118743},
	doi = {10.1051/0004-6361/201118743},
	language = {en},
	urldate = {2023-10-01},
	journal = {Astronomy \& Astrophysics},
	author = {Vilenius, E. and Kiss, C. and Mommert, M. and Müller, T. and Santos-Sanz, P. and Pal, A. and Stansberry, J. and Mueller, M. and Peixinho, N. and Fornasier, S. and Lellouch, E. and Delsanti, A. and Thirouin, A. and Ortiz, J. L. and Duffard, R. and Perna, D. and Szalai, N. and Protopapa, S. and Henry, F. and Hestroffer, D. and Rengel, M. and Dotto, E. and Hartogh, P.},
	month = may,
	year = {2012},
	pages = {A94},
}

@article{santos-sanz_tnos_2012,
	title = {"{TNOs} are {Cool}": {A} survey of the trans-{Neptunian} region. {IV}. {Size}/albedo characterization of 15 scattered disk and detached objects observed with {Herschel}-{PACS}},
	volume = {541},
	issn = {0004-6361},
	shorttitle = {"{TNOs} are {Cool}"},
	doi = {10.1051/0004-6361/201118541},
	journal = {Astronomy \& Astrophysics},
	author = {Santos-Sanz, P. and Lellouch, E. and Fornasier, S. and Kiss, C. and Pal, A. and Müller, T. G. and Vilenius, E. and Stansberry, J. and Mommert, M. and Delsanti, A. and Mueller, M. and Peixinho, N. and Henry, F. and Ortiz, J. L. and Thirouin, A. and Protopapa, S. and Duffard, R. and Szalai, N. and Lim, T. and Ejeta, C. and Hartogh, P. and Harris, A. W. and Rengel, M.},
	month = may,
	year = {2012},
	pages = {A92},
}

@article{vilenius_tnos_2014,
	title = {“{TNOs} are {Cool}”: {A} survey of the trans-{Neptunian} region - {X}. {Analysis} of classical {Kuiper} belt objects from {Herschel} and {Spitzer} observations},
	volume = {564},
	copyright = {© ESO, 2014},
	issn = {0004-6361, 1432-0746},
	shorttitle = {“{TNOs} are {Cool}”},
	url = {https://www.aanda.org/articles/aa/abs/2014/04/aa22416-13/aa22416-13.html},
	doi = {10.1051/0004-6361/201322416},
	language = {en},
	urldate = {2024-04-04},
	journal = {Astronomy \& Astrophysics},
	author = {Vilenius, E. and Kiss, C. and Müller, T. and Mommert, M. and Santos-Sanz, P. and Pál, A. and Stansberry, J. and Mueller, M. and Peixinho, N. and Lellouch, E. and Fornasier, S. and Delsanti, A. and Thirouin, A. and Ortiz, J. L. and Duffard, R. and Perna, D. and Henry, F.},
	month = apr,
	year = {2014},
	pages = {A35},
}

@incollection{muller_chapter_2020,
	title = {Chapter 7 - {Trans}-{Neptunian} objects and {Centaurs} at thermal wavelengths},
	isbn = {978-0-12-816490-7},
	url = {https://www.sciencedirect.com/science/article/pii/B9780128164907000072},
	urldate = {2024-04-17},
	booktitle = {The {Trans}-{Neptunian} {Solar} {System}},
	publisher = {Elsevier},
	author = {Müller, Thomas and Lellouch, Emmanuel and Fornasier, Sonia},
	editor = {Prialnik, Dina and Barucci, M. Antonietta and Young, Leslie A.},
	month = jan,
	year = {2020},
	doi = {10.1016/B978-0-12-816490-7.00007-2},
	pages = {153--181},
}

@article{harris_thermal_1998,
	title = {A {Thermal} {Model} for {Near}-{Earth} {Asteroids}},
	volume = {131},
	issn = {0019-1035},
	url = {https://www.sciencedirect.com/science/article/pii/S0019103597958656},
	doi = {10.1006/icar.1997.5865},
	number = {2},
	urldate = {2024-09-02},
	journal = {Icarus},
	author = {Harris, Alan W.},
	month = feb,
	year = {1998},
	pages = {291--301},
}

@article{delbo_thermal_2007,
	title = {Thermal inertia of near-{Earth} asteroids and implications for the magnitude of the {Yarkovsky} effect},
	volume = {190},
	issn = {0019-1035},
	url = {https://www.sciencedirect.com/science/article/pii/S0019103507001091},
	doi = {10.1016/j.icarus.2007.03.007},
	number = {1},
	urldate = {2024-09-03},
	journal = {Icarus},
	author = {Delbo, Marco and dell'Oro, Aldo and Harris, Alan W. and Mottola, Stefano and Mueller, Michael},
	month = sep,
	year = {2007},
	pages = {236--249},
}

@article{lellouch_tnos_2013,
	title = {“{TNOs} are {Cool}”: {A} survey of the trans-{Neptunian} region - {IX}. {Thermal} properties of {Kuiper} belt objects and {Centaurs} from combined {Herschel} and {Spitzer} observations},
	volume = {557},
	copyright = {© ESO, 2013},
	issn = {0004-6361, 1432-0746},
	shorttitle = {“{TNOs} are {Cool}”},
	url = {https://www.aanda.org/articles/aa/abs/2013/09/aa22047-13/aa22047-13.html},
	doi = {10.1051/0004-6361/201322047},
	language = {en},
	urldate = {2024-09-03},
	journal = {Astronomy \& Astrophysics},
	author = {Lellouch, E. and Santos-Sanz, P. and Lacerda, P. and Mommert, M. and Duffard, R. and Ortiz, J. L. and Müller, T. G. and Fornasier, S. and Stansberry, J. and Kiss, Cs and Vilenius, E. and Mueller, M. and Peixinho, N. and Moreno, R. and Groussin, O. and Delsanti, A. and Harris, A. W.},
	month = sep,
	year = {2013},
	note = {},
	pages = {A60},
}

@article{kiss_visible_2024,
	title = {The visible and thermal light curve of the large {Kuiper} belt object (50000) {Quaoar}},
	volume = {684},
	copyright = {© The Authors 2024},
	issn = {0004-6361, 1432-0746},
	url = {https://www.aanda.org/articles/aa/abs/2024/04/aa48054-23/aa48054-23.html},
	doi = {10.1051/0004-6361/202348054},
	language = {en},
	urldate = {2025-01-16},
	journal = {Astronomy \& Astrophysics},
	author = {Kiss, C. and Müller, T. G. and Marton, G. and Szakáts, R. and Pál, A. and Molnár, L. and Vilenius, E. and Rengel, M. and Ortiz, J. L. and Fernández-Valenzuela, E.},
	month = apr,
	year = {2024},
	note = {},
	pages = {A50},
}

@article{mommert_tnos_2012,
	title = {{TNOs} are cool: {A} survey of the trans-{Neptunian} region - {V}. {Physical} characterization of 18 {Plutinos} using {Herschel}-{PACS} observations},
	volume = {541},
	copyright = {© ESO, 2012},
	issn = {0004-6361, 1432-0746},
	shorttitle = {{TNOs} are cool},
	url = {https://www.aanda.org/articles/aa/abs/2012/05/aa18562-11/aa18562-11.html},
	doi = {10.1051/0004-6361/201118562},
	language = {en},
	urldate = {2025-01-16},
	journal = {Astronomy \& Astrophysics},
	author = {Mommert, M. and Harris, A. W. and Kiss, C. and Pál, A. and Santos-Sanz, P. and Stansberry, J. and Delsanti, A. and Vilenius, E. and Müller, T. G. and Peixinho, N. and Lellouch, E. and Szalai, N. and Henry, F. and Duffard, R. and Fornasier, S. and Hartogh, P. and Mueller, M. and Ortiz, J. L. and Protopapa, S. and Rengel, M. and Thirouin, A.},
	month = may,
	year = {2012},
	note = {},
	pages = {A93},
}

@article{santos-sanz_tnos_2017,
	title = {“{TNOs} are {Cool}”: {A} survey of the trans-{Neptunian} region - {XII}. {Thermal} light curves of {Haumea}, 2003 {VS2} and 2003 {AZ84} with {Herschel}/{PACS}},
	volume = {604},
	copyright = {© ESO, 2017},
	issn = {0004-6361, 1432-0746},
	shorttitle = {“{TNOs} are {Cool}”},
	url = {https://www.aanda.org/articles/aa/abs/2017/08/aa30354-16/aa30354-16.html},
	doi = {10.1051/0004-6361/201630354},
	language = {en},
	urldate = {2025-02-10},
	journal = {Astronomy \& Astrophysics},
	author = {Santos-Sanz, P. and Lellouch, E. and Groussin, O. and Lacerda, P. and Müller, T. G. and Ortiz, J. L. and Kiss, C. and Vilenius, E. and Stansberry, J. and Duffard, R. and Fornasier, S. and Jorda, L. and Thirouin, A.},
	month = aug,
	year = {2017},
	note = {},
	pages = {A95},
}

@article{fornasier_tnos_2013,
	title = {{TNOs} are {Cool}: {A} survey of the trans-{Neptunian} region - {VIII}. {Combined} {Herschel} {PACS} and {SPIRE} observations of nine bright targets at 70–500 μm},
	volume = {555},
	copyright = {© ESO, 2013},
	issn = {0004-6361, 1432-0746},
	shorttitle = {{TNOs} are {Cool}},
	url = {https://www.aanda.org/articles/aa/abs/2013/07/aa21329-13/aa21329-13.html},
	doi = {10.1051/0004-6361/201321329},
	language = {en},
	urldate = {2025-03-18},
	journal = {Astronomy \& Astrophysics},
	author = {Fornasier, S. and Lellouch, E. and Müller, T. and Santos-Sanz, P. and Panuzzo, P. and Kiss, C. and Lim, T. and Mommert, M. and Bockelée-Morvan, D. and Vilenius, E. and Stansberry, J. and Tozzi, G. P. and Mottola, S. and Delsanti, A. and Crovisier, J. and Duffard, R. and Henry, F. and Lacerda, P. and Barucci, A. and Gicquel, A.},
	month = jul,
	year = {2013},
	note = {},
	pages = {A15},
}

@article{farkas-takacs_tnos_2020,
	title = {“{TNOs} are {Cool}”: {A} survey of the trans-{Neptunian} region - {XV}. {Physical} characteristics of 23 resonant trans-{Neptunian} and scattered disk objects},
	volume = {638},
	copyright = {© ESO 2020},
	issn = {0004-6361, 1432-0746},
	shorttitle = {“{TNOs} are {Cool}”},
	url = {https://www.aanda.org/articles/aa/abs/2020/06/aa36183-19/aa36183-19.html},
	doi = {10.1051/0004-6361/201936183},
	language = {en},
	urldate = {2025-03-21},
	journal = {Astronomy \& Astrophysics},
	author = {Farkas-Takács, A. and Kiss, Cs and Vilenius, E. and Marton, G. and Müller, T. G. and Mommert, M. and Stansberry, J. and Lellouch, E. and Lacerda, P. and Pál, A.},
	month = jun,
	year = {2020},
	note = {},
	pages = {A23},
}

@article{lim_tnos_2010,
	title = {“{TNOs} are {Cool}”: {A} survey of the trans-{Neptunian} region - {III}. {Thermophysical} properties of 90482 {Orcus} and 136472 {Makemake}},
	volume = {518},
	copyright = {© ESO, 2010},
	issn = {0004-6361, 1432-0746},
	shorttitle = {“{TNOs} are {Cool}”},
	url = {https://www.aanda.org/articles/aa/abs/2010/10/aa14701-10/aa14701-10.html},
	doi = {10.1051/0004-6361/201014701},
	language = {en},
	urldate = {2025-03-21},
	journal = {Astronomy \& Astrophysics},
	author = {Lim, T. L. and Stansberry, J. and Müller, T. G. and Mueller, M. and Lellouch, E. and Kiss, C. and Santos-Sanz, P. and Vilenius, E. and Protopapa, S. and Moreno, R. and Delsanti, A. and Duffard, R. and Fornasier, S. and Groussin, O. and Harris, A. W. and Henry, F. and Horner, J. and Lacerda, P. and Mommert, M. and Ortiz, J. L. and Rengel, M. and Thirouin, A. and Trilling, D. and Barucci, A. and Crovisier, J. and Doressoundiram, A. and Dotto, E. and Buenestado, P. J. Gutiérrez and Hainaut, O. and Hartogh, P. and Hestroffer, D. and Kidger, M. and Lara, L. and Swinyard, B. M. and Thomas, N.},
	month = jul,
	year = {2010},
	note = {},
	pages = {L148},
}

@article{kiss_prominent_2024,
	title = {Prominent {Mid}-infrared {Excess} of the {Dwarf} {Planet} (136472) {Makemake} {Discovered} by {JWST}/{MIRI} {Indicates} {Ongoing} {Activity}},
	volume = {976},
	issn = {2041-8205},
	url = {https://dx.doi.org/10.3847/2041-8213/ad8dcb},
	doi = {10.3847/2041-8213/ad8dcb},
	language = {en},
	number = {1},
	urldate = {2025-05-06},
	journal = {The Astrophysical Journal Letters},
	author = {Kiss, Csaba and Müller, Thomas G. and Farkas-Takács, Anikó and Moór, Attila and Protopapa, Silvia and Parker, Alex H. and Santos-Sanz, Pablo and Ortiz, Jose Luis and Holler, Bryan J. and Wong, Ian and Stansberry, John and Fernández-Valenzuela, Estela and Glein, Christopher R. and Lellouch, Emmanuel and Vilenius, Esa and Kalup, Csilla E. and Regály, Zsolt and Szakáts, Róbert and Marton, Gábor and Pál, András and Szabó, Gyula M.},
	month = nov,
	year = {2024},
	note = {},
	pages = {L9},
}

@article{nesvorny_trans-neptunian_2019,
	title = {Trans-{Neptunian} binaries as evidence for planetesimal formation by the streaming instability},
	volume = {3},
	copyright = {2019 The Author(s), under exclusive licence to Springer Nature Limited},
	issn = {2397-3366},
	url = {https://www.nature.com/articles/s41550-019-0806-z},
	doi = {10.1038/s41550-019-0806-z},
	language = {en},
	number = {9},
	urldate = {2024-01-16},
	journal = {Nature Astronomy},
	author = {Nesvorný, David and Li, Rixin and Youdin, Andrew N. and Simon, Jacob B. and Grundy, William M.},
	month = sep,
	year = {2019},
	pages = {808--812},
}

@article{fraser_all_2017,
	title = {All planetesimals born near the {Kuiper} belt formed as binaries},
	volume = {1},
	copyright = {2017 Macmillan Publishers Limited, part of Springer Nature.},
	issn = {2397-3366},
	url = {https://www.nature.com/articles/s41550-017-0088},
	doi = {10.1038/s41550-017-0088},
	language = {en},
	number = {4},
	urldate = {2024-01-29},
	journal = {Nature Astronomy},
	author = {Fraser, Wesley C. and Bannister, Michele T. and Pike, Rosemary E. and Marsset, Michael and Schwamb, Megan E. and Kavelaars, J. J. and Lacerda, Pedro and Nesvorný, David and Volk, Kathryn and Delsanti, Audrey and Benecchi, Susan and Lehner, Matthew J. and Noll, Keith and Gladman, Brett and Petit, Jean-Marc and Gwyn, Stephen and Chen, Ying-Tung and Wang, Shiang-Yu and Alexandersen, Mike and Burdullis, Todd and Sheppard, Scott and Trujillo, Chad},
	month = apr,
	year = {2017},
	pages = {1--6},
}

@article{nesvorny_binary_2019,
	title = {Binary survival in the outer solar system},
	volume = {331},
	issn = {0019-1035},
	url = {https://www.sciencedirect.com/science/article/pii/S0019103518306900},
	doi = {10.1016/j.icarus.2019.04.030},
	urldate = {2024-01-30},
	journal = {Icarus},
	author = {Nesvorný, David and Vokrouhlický, David},
	month = oct,
	year = {2019},
	pages = {49--61},
}

@article{grundy_mutual_2019,
	series = {The {Trans}-{Neptunian} {Solar} {System}},
	title = {Mutual orbit orientations of transneptunian binaries},
	volume = {334},
	issn = {0019-1035},
	doi = {10.1016/j.icarus.2019.03.035},
	urldate = {2024-01-30},
	journal = {Icarus},
	author = {Grundy, W. M. and Noll, K. S. and Roe, H. G. and Buie, M. W. and Porter, S. B. and Parker, A. H. and Nesvorný, D. and Levison, H. F. and Benecchi, S. D. and Stephens, D. C. and Trujillo, C. A.},
	year = {2019},
	pages = {62--78},
}

@incollection{noll_chapter_2020,
	title = {Chapter 9 - {Trans}-{Neptunian} {Binaries} (2018)},
	isbn = {978-0-12-816490-7},
	urldate = {2023-10-02},
	booktitle = {The {Trans}-{Neptunian} {Solar} {System}},
	publisher = {Elsevier},
	author = {Noll, Keith S. and Grundy, William M. and Nesvorný, David and Thirouin, Audrey},
	editor = {Prialnik, Dina and Barucci, M. Antonietta and Young, Leslie A.},
	month = jan,
	year = {2020},
	doi = {10.1016/B978-0-12-816490-7.00009-6},
	pages = {205--224},
}

@article{porter_detection_2024,
	title = {Detection of {Close} {Kuiper} {Belt} {Binaries} with {HST} {WFC3}},
	volume = {5},
	issn = {2632-3338},
	url = {https://iopscience.iop.org/article/10.3847/PSJ/ad3f19/meta},
	doi = {10.3847/PSJ/ad3f19},
	language = {en},
	number = {6},
	urldate = {2025-01-20},
	journal = {The Planetary Science Journal},
	author = {Porter, Simon B. and Benecchi, Susan D. and Verbiscer, Anne J. and Grundy, W. M. and Noll, Keith S. and Parker, Alex H.},
	month = jun,
	year = {2024},
	note = {},
	pages = {143},
}

@article{parker_characterization_2011,
	title = {Characterization of seven ultra-wide trans-{Neptunian} binaries},
	volume = {743},
	issn = {0004-637X},
	url = {https://dx.doi.org/10.1088/0004-637X/743/1/1},
	doi = {10.1088/0004-637X/743/1/1},
	language = {en},
	number = {1},
	urldate = {2025-01-27},
	journal = {The Astrophysical Journal},
	author = {Parker, Alex H. and Kavelaars, J. J. and Petit, Jean-Marc and Jones, Lynne and Gladman, Brett and Parker, Joel},
	month = nov,
	year = {2011},
	note = {},
	pages = {1},
}

@article{grundy_five_2011,
	title = {Five new and three improved mutual orbits of transneptunian binaries},
	volume = {213},
	issn = {0019-1035},
	doi = {10.1016/j.icarus.2011.03.012},
	number = {2},
	urldate = {2025-02-03},
	journal = {Icarus},
	author = {Grundy, W. M. and Noll, K. S. and Nimmo, F. and Roe, H. G. and Buie, M. W. and Porter, S. B. and Benecchi, S. D. and Stephens, D. C. and Levison, H. F. and Stansberry, J. A.},
	year = {2011},
	pages = {678--692},
}

@article{lyra_where_2025,
	title = {Where are the missing {Kuiper} {Belt} binaries?},
	volume = {442},
	issn = {0019-1035},
	url = {https://www.sciencedirect.com/science/article/pii/S0019103525002854},
	doi = {10.1016/j.icarus.2025.116737},
	urldate = {2025-08-25},
	journal = {Icarus},
	author = {Lyra, Wladimir},
	month = dec,
	year = {2025},
	pages = {116737},
}

@article{thirouin2010short,
  title={Short-term variability of a sample of 29 trans-Neptunian objects and Centaurs},
  author={Thirouin, A and Ortiz, JL and Duffard, R and Santos-Sanz, P and Aceituno, FJ and Morales, N},
  journal={Astronomy \& Astrophysics},
  volume={522},
  pages={A93},
  year={2010},
  
}

@article{ortiz2003study,
  title={A study of short term rotational variability in TNOs and Centaurs from Sierra Nevada Observatory},
  author={Ortiz, JL and Guti{\'e}rrez, PJ and Casanova, V and Sota, A},
  journal={Astronomy \& Astrophysics},
  volume={407},
  number={3},
  pages={1149--1155},
  year={2003},
  
}

@article{porter_kctf_2012,
	title = {{KCTF} evolution of trans-neptunian binaries: {Connecting} formation to observation},
	volume = {220},
	issn = {0019-1035},
	shorttitle = {{KCTF} evolution of trans-neptunian binaries},
	url = {https://www.sciencedirect.com/science/article/pii/S0019103512002631},
	doi = {10.1016/j.icarus.2012.06.034},
	number = {2},
	urldate = {2025-08-29},
	journal = {Icarus},
	author = {Porter, Simon B. and Grundy, William M.},
	month = aug,
	year = {2012},
	pages = {947--957},
}


\begin{appendix} 
\section{Robustness analysis with respect to model assumptions} \label{ap:robustness}
\subsection{Effects of emissivity}
\begin{figure}[h!]
\centering

\includegraphics[width=\linewidth]{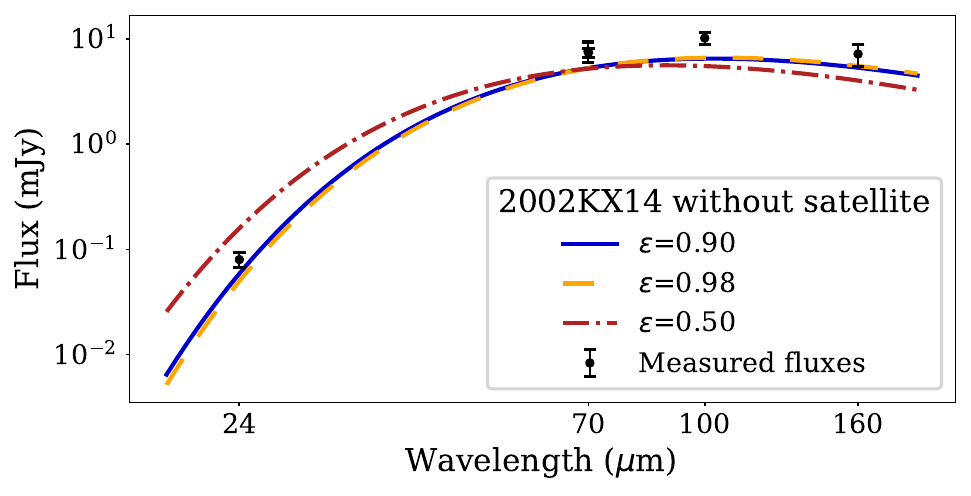}

\caption{\kxlong{} thermal SED simulations assuming no satellite and different emissivities. We considered the $D_{\text{eq}}$ and $H_V$ values in Table \ref{tab:merged_table} and $\eta = 1.8$. No value of $\epsilon$ delivers a good fit to the data.}

\label{fig:emissivity_plot}
\end{figure}

Emissivity is a measure of a material's surface efficiency in emitting thermal radiation compared to a perfect blackbody. By definition, it is bounded in the [0, 1] range. The choice of an emissivity $\epsilon = 0.9$ for the "TNOs are Cool" works is based on laboratory measurements of silicate powder \citep{hovis1966infrared}, which is thought to resemble the properties of small Solar System bodies \citep{vilenius_tnos_2012}. It is reasonable to question whether a different value of $\epsilon$ might explain the discrepant occultation and thermal measurements without the need to invoke a satellite. To test this possibility, we take \kx{} as an example. In Figure \ref{fig:emissivity_plot} we plot the spectral energy distribution (SED) of \kx{} simulated under the NEATM considering a single body with the observed absolute magnitude and area-equivalent diameter. For a reasonable choice of $\eta=1.8$ (see Figure \ref{fig:cornerplots_rest} and \cite{vilenius_tnos_2012}), we considered the standard emissivity $\epsilon = 0.9$, a high emissivity of $\epsilon = 0.98$ typical of water ice, and a very damped emissivity of $\epsilon = 0.5$. None of these choices can explain the flux values observed in the thermal measurements. We conclude that the observed thermal excesses are not attributable to emissivity values different from $\epsilon = 0.9$. 

\subsection{Effects of non-spherical shapes}

While the standard NEATM assumes a spherical shape, the observed occultation projections of our targets indicate significant oblateness. It is therefore necessary to test whether accounting for non-spherical geometries could resolve the inconsistencies between the thermal and occultation data, thereby rendering the unresolved satellite hypothesis unnecessary.

We performed simulations with a thermal model identical to the NEATM, but assuming an ellipsoidal shape with semi-axes $a\ge b \ge c$, where the minor axis is coincident with the spin axis. These simulations are performed for the objects  for which we infer the presence of previously undetected satellites: \tclong{}, \kxlong{}, and \manilong{}. For each object we considered the $\eta$ value fitted within the NEATM in the "TNOs are Cool" papers \citep{vilenius_tnos_2012, fornasier_tnos_2013}. We considered only ellipsoids compatible with the observed light curve amplitudes and projected elliptical limbs (with semi-axes $a', b'$) in occultations. To take this into account, we generated  random ellipsoids for each object. We chose random spin pole orientations. The semi-axes values are chosen randomly within reasonable bounds. For \kx{} and \mani{} we took into account that a spheroidal shape (where $a = b = a'$) is likely \citep{rizos2025trans, rommel2023large}:
\begin{itemize}
    \item \tc{}: $a \in [250, 300]$~km, $b\in [a/2, a]$~km, $c\in [b/2, b]$~km. 
    \item \kx{}: $a \in [230, 250]$~km, $b\in [0.9a, a]$~km, $c\in [b/2, b]$~km. 
    \item \mani{}: $a \in [400, 425]$~km, $b\in [0.9a, a]$~km, $c\in [b/2, b]$~km.
\end{itemize}

\begin{figure*}[ht!]
\centering

\begin{subfigure}{0.48\textwidth}
    \centering
    \includegraphics[width=\linewidth]{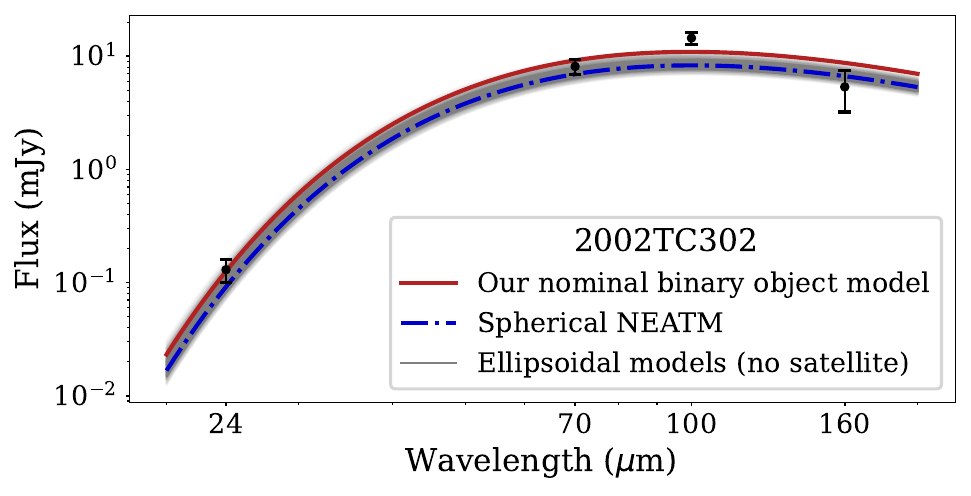}
\end{subfigure}
\hfill
\begin{subfigure}{0.48\textwidth}
    \centering
    \includegraphics[width=\linewidth]{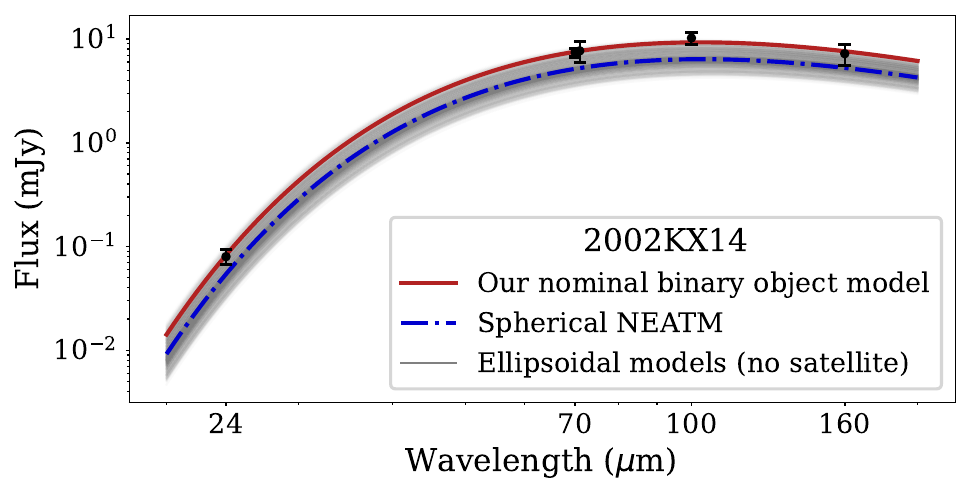}

\end{subfigure}

\hspace*{\fill}
\begin{subfigure}{0.48\textwidth}
    \centering
    \includegraphics[width=\linewidth]{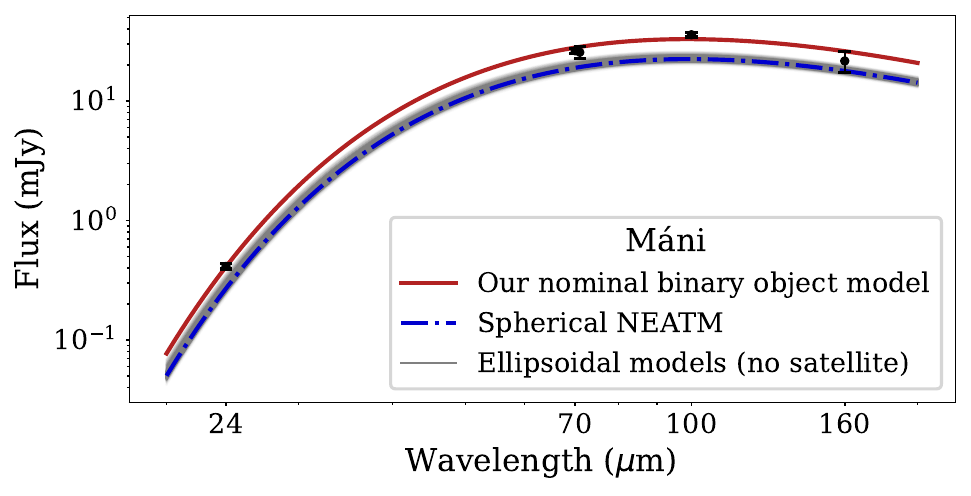}

\end{subfigure}
\hspace*{\fill}

\caption{Ellipsoidal models vs. our nominal binary object models for the targets for which we infer the presence of a satellite. The SED for the ellipsoidal models tend to underestimate the measured thermal fluxes, while a binary model provides a better fit to the data.  The single-object spherical NEATM with the parameters form Table \ref{tab:merged_table} is also plotted for reference.}
\label{fig:ellipsoidal}
\end{figure*}

We excluded ellipsoidal models whose light curve amplitudes according to \cite[Eq. 5, p. 426]{binzel_asteroids_1989} were inconsistent with the values in Table \ref{tab:merged_table} within $1\sigma$. Additionally, we discarded models where the projected elliptical limb semi-axes at the occultation epoch deviated by more than $1\sigma$ from the fitted values reported  in \cite{ortiz2020large}, \cite{rizos2025trans} and \cite{rommel2023large} respectively.  We repeated the generation of random ellipsoids iteratively until we had 1000 valid models for each object, which are shown in Figure \ref{fig:ellipsoidal}.

For each valid model, we calculated the simulated-to-measured flux ratios (Table \ref{tab:flux_ratios}). The conclusion is that single-object ellipsoidal models generally tend to underestimate the observed flux. A small subset of models for \tc{} and \kx{} could reproduce the observations. However, such configurations are statistically improbable under our assumption of random pole orientations.

Crucially, this analysis demonstrates that our methodology is not biased toward satellite detection. Departing from a spherical approximation introduces scatter but does not result in systematically higher fluxes. As shown in Figure \ref{fig:ellipsoidal}, the valid ellipsoidal models yield an ensemble of SEDs centered around the single-body NEATM prediction, confirming that the observed flux excesses cannot be systematically attributed to non-spherical shape effects.

\begin{table}
\caption{Simulated-to-measured flux ratios for single-object ellipsoidal models.}
\label{tab:flux_ratios}
\renewcommand{\arraystretch}{1.3} 
\centering
\begin{tabular}{p{1 cm}p{1 cm}p{1 cm}p{1.2 cm}p{1.2 cm}p{1.2 cm}}
\hline \hline
Object & 24 \micron{} & 71 \micron{} & 70 \micron{} & 100 \micron{} & 160 \micron{}\\
\hline
\tc{} & $0.79^{+0.09}_{-0.16}$ & - & $0.89^{+0.07}_{-0.11}$ & $0.59^{+0.04}_{-0.06}$ & $1.27^{+0.08}_{-0.13}$\\
\kx{} & $0.69^{+0.07}_{-0.29}$ & $0.70^{+0.09}_{-0.21}$  & $0.72^{+0.09}_{-0.21}$ & $0.65^{+0.08}_{-0.17}$ & $0.76^{+0.13}_{-0.15}$\\
\mani{} & $0.68^{+0.04}_{-0.06}$ & $0.78^{+0.03}_{-0.04}$ & $0.73^{+0.03}_{-0.04}$ & $0.64^{+0.03}_{-0.03}$ & $0.83^{+0.03}_{-0.04}$\\
\hline
\bottomrule
\end{tabular}
\tablefoot{We report the mean values across the 1000 accepted models as the nominal values and we compute the 68\% highest density intervals to provide errorbars.}
\end{table}

\section{Data-against-model plots} \label{ap:results}

\begin{figure*}
\centering

\begin{subfigure}{0.45\textwidth}
\includegraphics[width=\linewidth]{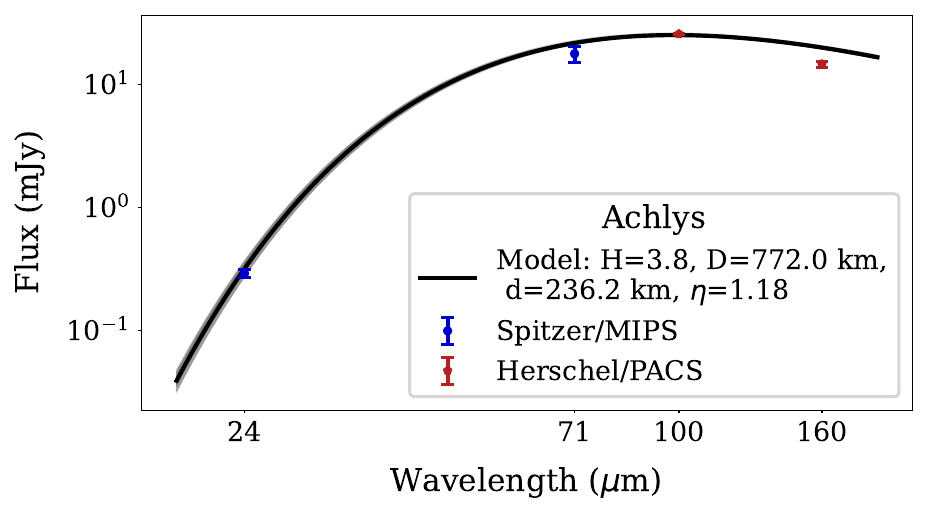}

\end{subfigure}
\hfill
\begin{subfigure}{0.45\textwidth}
\includegraphics[width=\linewidth]{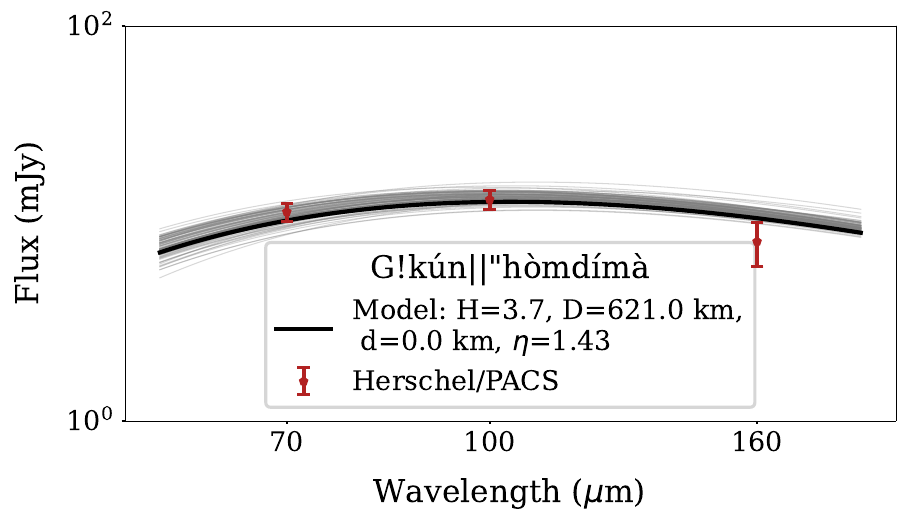}

\end{subfigure}

\begin{subfigure}{0.45\textwidth}
\includegraphics[width=\linewidth]{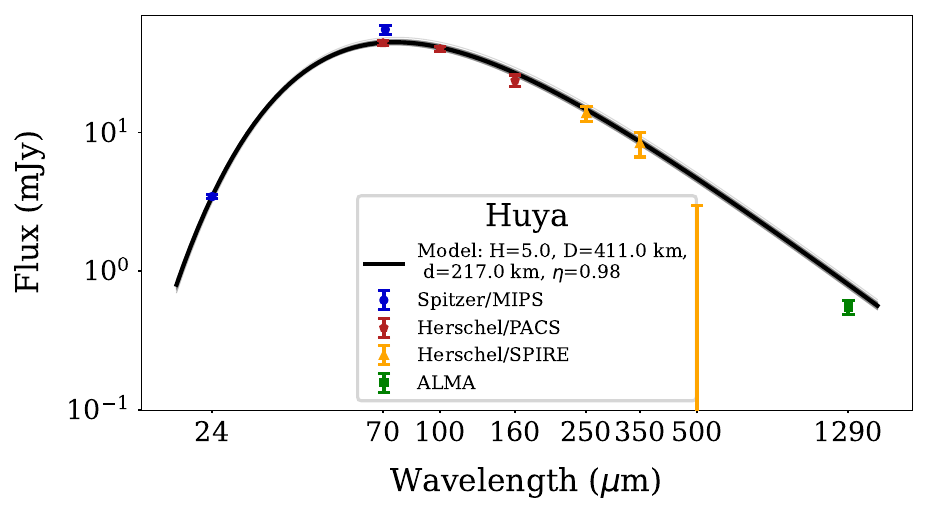}

\end{subfigure}
\hfill
\begin{subfigure}{0.45\textwidth}
\includegraphics[width=\linewidth]{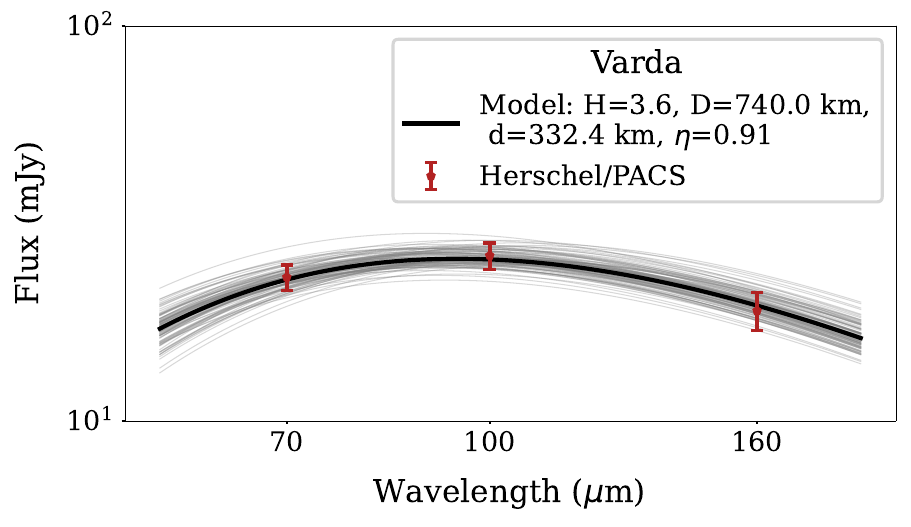}

\end{subfigure}

\caption{Modelled thermal emission for validation targets. See Appendix~\ref{ap:results} for details.}
\label{fig:emission_group1}
\end{figure*}

\begin{figure*}
\centering

\begin{subfigure}{0.45\textwidth}
\includegraphics[width=\linewidth]{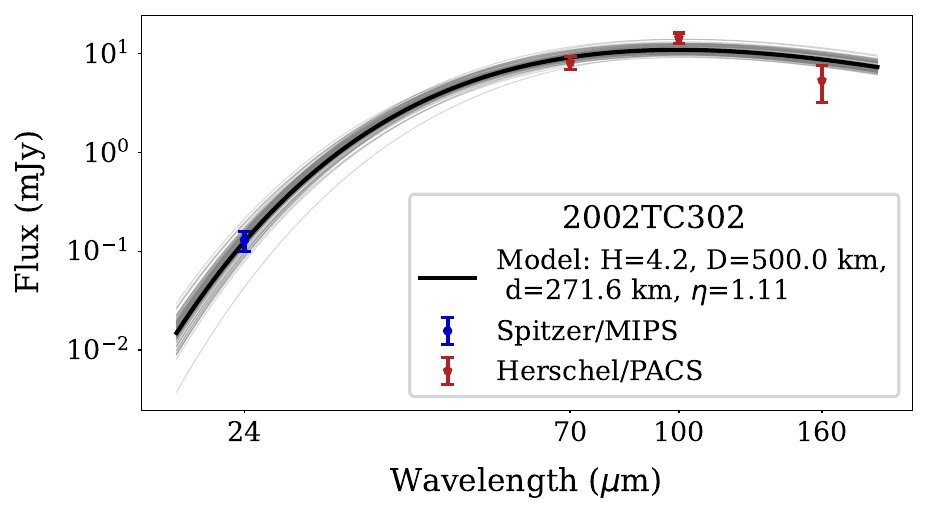}

\end{subfigure}
\hfill
\begin{subfigure}{0.45\textwidth}
\includegraphics[width=\linewidth]{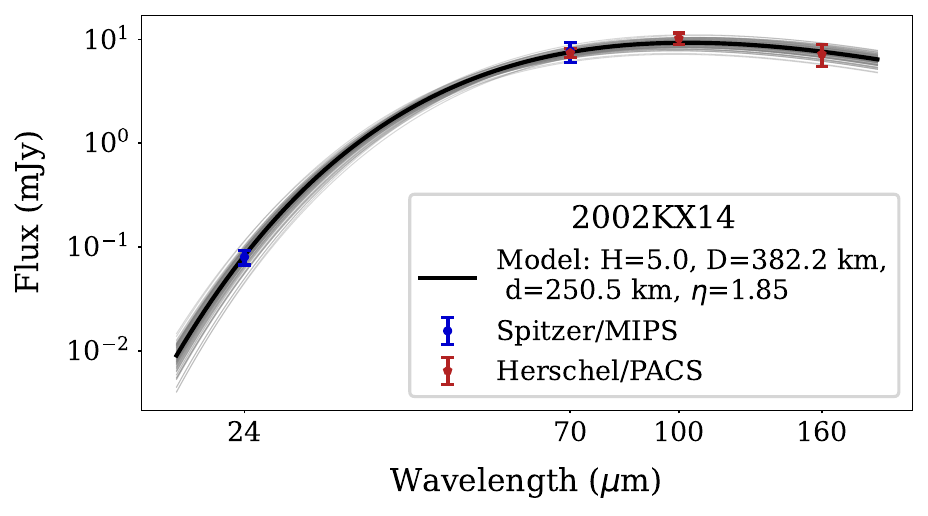}

\end{subfigure}

\begin{subfigure}{0.45\textwidth}
\includegraphics[width=\linewidth]{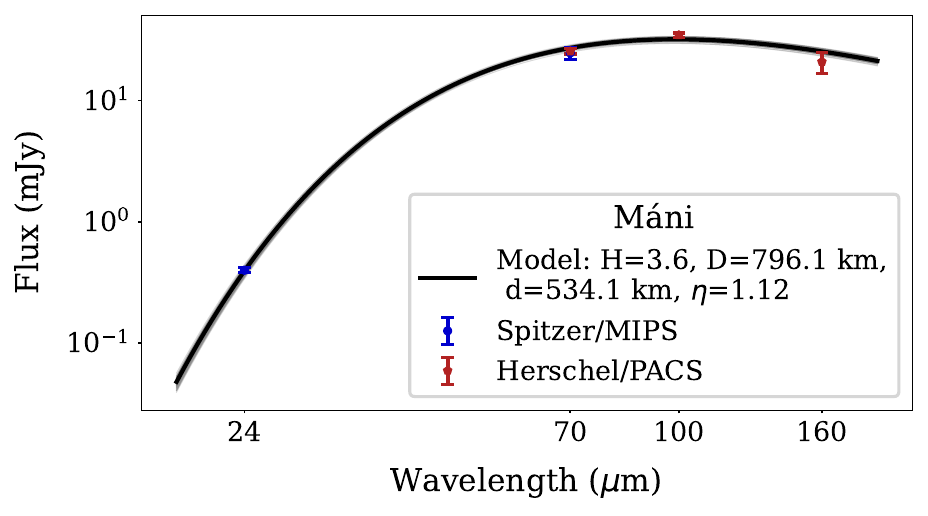}

\end{subfigure}
\hfill
\begin{subfigure}{0.45\textwidth}
\includegraphics[width=\linewidth]{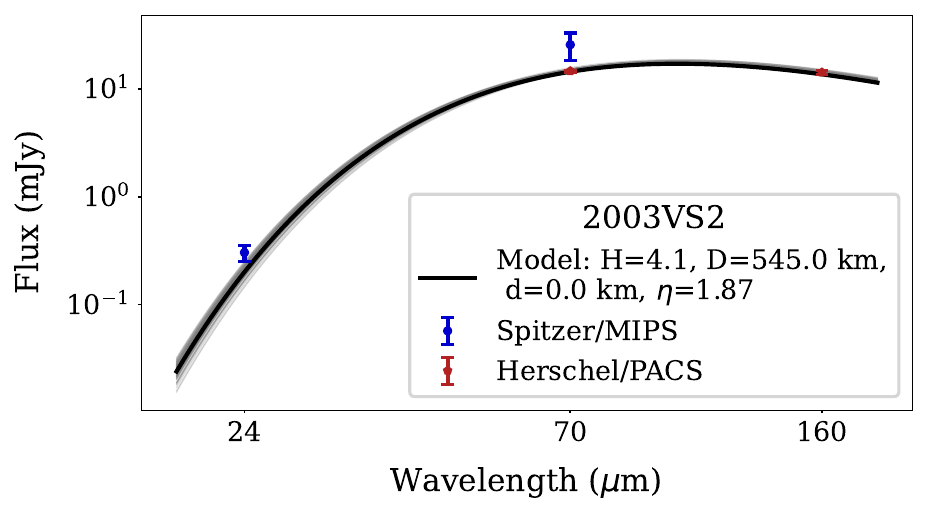}

\end{subfigure}

\begin{subfigure}{0.45\textwidth}
\includegraphics[width=\linewidth]{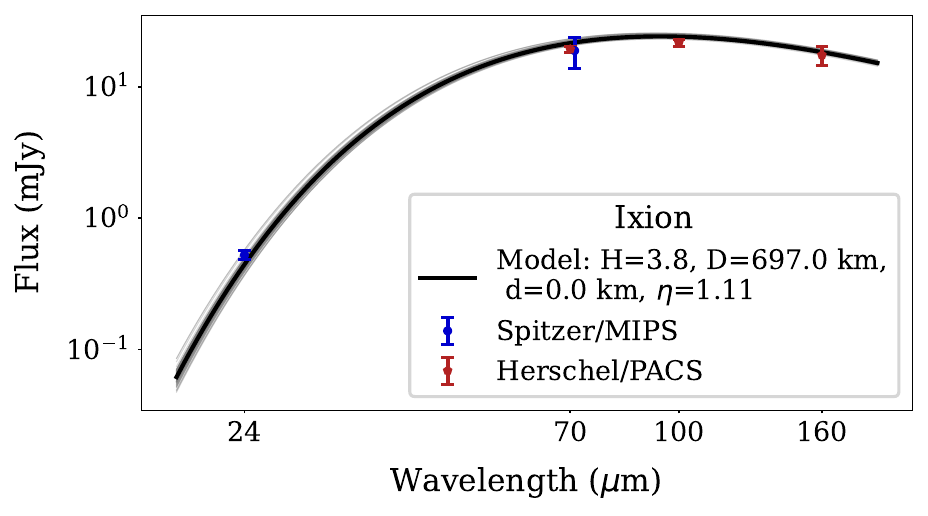}

\end{subfigure}
\hfill
\begin{subfigure}{0.45\textwidth}
\includegraphics[width=\linewidth]{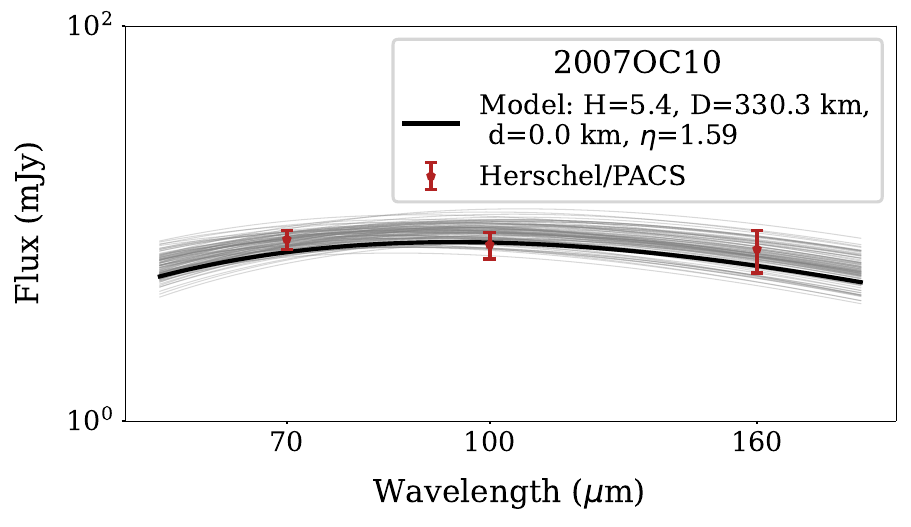}

\end{subfigure}

\caption{Modelled thermal emission for non-validation targets. See Appendix~\ref{ap:results} for details.}
\label{fig:emission_group2}
\end{figure*}

In this appendix section we show multi-panel Figures \ref{fig:emission_group1} and \ref{fig:emission_group2}. Each panel corresponds to one target. We plot our modelled SED for the system at thermal wavelengths in a thick black line, for the nominal parameter values shown in the legend, versus the measured thermal fluxes from the "TNOs are Cool" project. Thermal fluxes measured with different instruments are also measured at different epochs and different geometric configurations. We normalize every flux value to the heliocentric distance and target-observer distance from the \textit{Herschel}/PACS observing epoch, which is also the configuration for which the thermal SED is plotted. We also plot 100 randomly selected samples from our posterior MCMC estimates, displayed in thin grey lines, representing the uncertainties in our parameter estimates.

\end{appendix} 

\end{document}